\documentclass[epj]{svjour}
\usepackage{graphics}
\usepackage{widetext}
\pdfoutput=1 

\usepackage{color}
\definecolor{brown}{cmyk}{0, 0.8, 1, 0.6}
\definecolor{orange}{rgb}{1,0.5,0}

\usepackage{amsmath}
\usepackage{amssymb}
\usepackage{latexsym}
\usepackage{color}
\usepackage{bm}
\usepackage{multirow}
\usepackage{cancel}
\usepackage{hyperref}

\def\beq{\begin{eqnarray}}
\def\eeq{\end{eqnarray}}
\def\beq{\begin{equation}}
\def\eeq{\end{equation}}
\def\be{\begin{eqnarray}}
\def\ed{\end{eqnarray}}

\def\gaa{\mathrel{\raise.3ex\hbox{$>$\kern-.75em\lower1ex\hbox{$\sim$}}}}
\def\la{\mathrel{\raise.3ex\hbox{$<$\kern-.75em\lower1ex\hbox{$\sim$}}}}
\newcommand{\ba}{\begin{array}}
\newcommand{\ea}{\end{array}}
\newcommand{\besub}{\begin{subequations}}
\newcommand{\eesub}{\end{subequations}}
\newcommand{\ee}{\end{equation}}
\newcommand{\bea}{\begin{eqnarray}}
\newcommand{\eea}{\end{eqnarray}}

\begin{document}
\title{\Large Extended Higgs sector of 2HDM with real singlet facing LHC data}
\author{A.~Arhrib\inst{1} \and R.~Benbrik\inst{2,3} \and M.~EL Kacimi\inst{3} \and L.~Rahili \inst{3,4} \and S.~Semlali\inst{3} 
}
%
%
\institute{Faculty of Sciences and techniques, Abdelmalek Essaadi University, B.P. 416. Tanger, Morocco \and
MSISM Team, Facult\'e Polydisciplinaire de Safi, Sidi Bouzid, B.P. 4162,  Safi, Morocco \and
LPHEA, Faculty of Science Semlalia, Cadi Ayyad University, P.O.B. 2390 Marrakech, Morocco \and
EPTHE, Faculty of Sciences, Ibn Zohr University, P.O.B. 8106 Agadir, Morocco
}
\date{Received: date / Revised version: date}
%
\abstract{
We study the phenomenology of the two Higgs doublet model with a real singlet scalar $S$ (2HDMS). The model 
predicts three CP-even Higgses $h_{1,2,3}$, one CP-odd $A^0$ and a pair of charged Higgs.
We discuss the consistency of the 2HDMS with theoretical as well as with all available experimental data.
In contrast with previous studies, we focus on the scenario where $h_2$ is the Standard Model (SM) 125 GeV Higgs, 
while $h_1$ is lighter than $h_2$ which may open a window for Higgs to Higgs decays. 
We perform an extensive scan into the parameter space of 2HDMS of type I and explore the effect of the singlet-doublet admixture. 
We found that a large singlet-doublet admixture is still compatible with the recent Higgs data from LHC. 
Moreover, we show that $h_1$ could be quasi-fermiophobic and would decay dominantly into two photons.
We also study in details the consistency of the non-detected decay of $h_2\to h_1 h_1$ with LHC data followed by 
$h_1\to \gamma \gamma$ which leads to four photons final state at  LHC: $pp\to h_2\to h_1 h_1\to 4\gamma$.  
Using the results of null searches of multi-photons carried by the ATLAS collaboration, we have found that a large area of the parameter space is still allowed.  We also demonstrate that various neutral Higgs of the 2HDMS could have several exotic 
decays.}

\maketitle

\section{Introduction}
\label{intro}
A Higgs-like particle has been discovered in the first run of Large Hadron Collider (LHC) with 7 and 8 TeV energy in 2012 \cite{atlasdiscovery,cmsdiscovery} and some of its properties, such as its decay to some Standard Model (SM) particles, 
have been established. During the second run of LHC with 13 TeV center of mass energy, some Higgs-like decay measurements get improved and new observable such as $pp\to t\bar t h$ was performed \cite{Sirunyan:2018hoz,Aaboud:2018urx}.  \\

The aforementioned Higgs-like properties established so far will be further improved by the High Luminosity program of  the future LHC run (HL-LHC). 
At the HL-LHC, one can pin down the uncertainties on the Higgs-like couplings to a few percent level in some cases \cite{accuracy1} and provide indirect hints to the awaited new physics. Moreover, in the clean environment of the $e^+e^-$ collider such as ILC and CEPC, which can act like a Higgs factory,  one can improve the Higgs-like properties \cite{accuracy2,accuracy3}. 

Although all data collected by LHC seems to indicate that the Higgs-like particle is in perfect agreement with the SM predictions, 
there are theoretical as well as experimental hints that indicate that the SM is only an effective field theory of a more fundamental one. 
One common feature of those Beyond SM (BSM) theories is an extended Higgs sector with an extra singlet,  doublet,  and/or triplet. 
Most of the higher Higgs representations with an extra doublet/singlet predict in 
their spectrum extra neutral and/or charged Higgs states. 
Discovery of another Higgs state or more would be seen as a clear evidence of an 
extended Higgs sector and a departure from the SM.

Following the discovery of a Higgs-like particle, there has been a large amount of 
works dedicated to extending the SM Higgs sector 
by extra Higgs fields. Among the simplest one we mention:

A real singlet scalar that has a mixing with the SM Higgs boson \cite{Robens:2015gla},
the popular Two Higgs Doublet Model (2HDM) with or without natural flavor conservation  \cite{Robens:2015gla},
the popular two Higgs Doublet Model (2HDM) with or without natural flavor 
conservation \cite{Bernon:2015qea} and the inert Higgs model that 
provides dark matter candidate \cite{Ma:2006km,Arhrib:2013ela}. 
Recently, there have been also phenomenological studies that extend the 2HDM
with an additional real gauge-singlet scalar which act as dark matter 
candidate \cite{Drozd:2014yla,Grzadkowski:2009iz}. 
One can also extend the 2HDM by adding a real scalar singlet with non-vanishing expectation value that can mix with the doublets 
\cite{Chen:2013jvg,Muhlleitner:2016mzt}, a model which we call 2HDMS. 
In the two variants of the 2HDMS, the scalar spectrum is richer than the traditional 2HDM, which imply an interesting phenomenology at colliders including but not restricted to scalar-to-scalar decays, exotic decays and fermiophobic scenarios which are precluded in the SM and occurs hardly in the 2HDM.  The model can easily accommodate a SM-like Higgs Boson that easily satisfies all the constraints from LHC measurements. Despite the existence of mixing among CP-even mass eigenstates, which would modify the SM-like Higgs couplings to fermions and gauge bosons, constraints from signal strength measurements can be easily satisfied (within the present range of systematic and statistical errors). 

In the 2HDMS, the scalar spectrum contains 3 CP-even states $h_{1,2,3}$, one CP-odd $A$  and a pair of charged Higgs $H^\pm$.  $h_{1,2,3}$ are admixtures of doublet and singlet components while $A$ and $H^\pm$ are pure doublet Higgs.
A comprehensive analysis of the 2HDMS has been carried by the authors of Ref. \cite{Muhlleitner:2016mzt} 
assuming that the lightest CP-even scalar ($h_1$) is the SM-like Higgs boson. 
In this scenario, and while satisfying all theoretical and experimental constraints, 
a large singlet-doublet admixture is still compatible with LEP and LHC data \cite{Muhlleitner:2016mzt}.
In the present study, we would like to do a comprehensive study for the case where $h_2$ is the 125 GeV Higgs while $h_1$ is lighter than $h_2$ which open a window for Higgs to Higgs decay such as $h_{2,3}\to h_1h_1, AZ, W^\pm H^\mp$ and also $h_3\to h_1h_2, h_2h_2,$ which are still compatible with Higgs data.

The paper is organized as follow: section 2 is devoted to the 2HDMS and its parametrization, in section 3 we review the theoretical and experimental constraints that the 2HDMS is subject to. We present our numerical result in section 4 and conclude in section 5. Several technical details are given in the Appendix.

\section{The 2HDM with a real singlet field: 2HDMS}
In this section, we present a review of the 2HDMS. We discuss the scalar potential and derive the spectrum and the parametrization of the model. We also present the Yukawa textures and discuss the natural flavor conservation of the model. 
Couplings of the Higgs bosons to gauge bosons are also shown and their sum rules are discussed.

\subsection{The Higgs potential}
\label{sec:thehiggspot}
The scalar sector of 2HDMS consists of two weak isospin doublets 
$H_{i}$ (i = 1,2), with hypercharge $Y= 1$ and a real singlet field with hypercharge $Y= 0$ which are given by
\begin{eqnarray}
H_{i} & = \left(
                    \begin{array}{c}
                    \phi_i^\pm \\
                    \frac{1}{\sqrt{2}}(v_i + \phi_i + i \chi_i) \\
                    \end{array}
                    \right)~~{\rm and}~~S =
                  \frac{1}{\sqrt{2}}(v_s + \phi_s)
\end{eqnarray}

The most general renormalizable scalar potential for the model that respect 
$SU(2)_L\otimes U(1)_Y$ gauge symmetry has the following form:

\begin{widetext}
\begin{eqnarray}
V(H_1,H_2,S) &=&
m^2_{11}\, H_1^\dagger H_1
+ m^2_{22}\, H_2^\dagger H_2 - \mu^2\, \left(H_1^\dagger H_2 + H_2^\dagger H_1\right) + \frac{1}{2}m^2_S S^2 \nonumber\\
&+& \frac{\lambda_1}{2} \left( H_1^\dagger H_1 \right)^2
+ \frac{\lambda_2}{2} \left( H_2^\dagger H_2 \right)^2 + \lambda_3\, H_1^\dagger H_1\, H_2^\dagger H_2 + \lambda_4\, H_1^\dagger H_2\, H_2^\dagger H_1\nonumber\\
&+&\frac{\lambda_5}{2} \left[\left( H_1^\dagger H_2 \right)^2+ \left( H_2^\dagger H_1 \right)^2 \right] +
\frac{\lambda_6}{8} S^4 +\frac{1}{2}[\lambda_7 H_1^\dagger H_1 +\lambda_8 H_2^\dagger H_2 ] S^2
\label{eq:Vpot}
\end{eqnarray}
\end{widetext}
where $m_{11}^2, m_{22}^2$ and $m_{S}^2$ are masses terms.
By hermiticity of the scalar potential  $\lambda_{1,2,3,4,6,7,8}$ are dimensionless real 
parameters while $\lambda_5$ and $\mu^2$ can be complex to allow CP violation in the scalar sector.
In the present study, we assume that all scalar parameters are real. Therefore, the only source of CP violation is in the 
Cabbibo-Kobayashi-Maskawa matrix.
We remind here that we allow a  dimension 2 term  $\mu^2$ which break softly $Z_2$ symmetry.
This discrete $Z_2$ symmetry is usually imposed in order to avoid Flavor 
Changing Neutral Current (FCNC) at tree level in the Yukawa sector.  

\noindent
Assuming that spontaneous electroweak symmetry breaking (EWSB) is taking place at some 
electrically neutral point in the field space, and denoting the corresponding VEVs by
\begin{eqnarray} 
\hspace{-0.2cm}                   
\langle H_{1} \rangle &=\frac{1}{\sqrt{2}}\left(
\begin{array}{c}
0 \\
v_1 \\
\end{array}
\right),\langle H_{2} \rangle =\frac{1}{\sqrt{2}}\left(
\begin{array}{c}
0 \\
v_2\\
\end{array}
\right)~{\rm and}~\langle S \rangle =\frac{1}{\sqrt{2}}v_s
\label{eq:VEVs}
\end{eqnarray}

The parameters $m_{11}^2, m_{22}^2$ and $m_{S}^2$ can be eliminated  by the minimization conditions of the potential Eq.(\ref{eq:Vpot}): 
\begin{eqnarray}
m_{11}^2 &=& \mu^2 t_\beta - \frac{1}{2}\lambda_1v^2 c^2_{\beta} - \frac{1}{2}\lambda_{345}v^2 s^2_{\beta} - \frac{1}{4}\lambda_7 v_S^2 \label{eq:ewsb1}\\
m_{22}^2 &=& \mu^2 t_\beta^{-1} - \frac{1}{2}\lambda_1v^2 s^2_{\beta}- \frac{1}{2}\lambda_{345}v^2 c^2_{\beta} - \frac{1}{4}\lambda_8 v_S^2  \label{eq:ewsb2}\\
m_{S}^2 &=& - \frac{1}{2}\lambda_7 v^2 c^2_{\beta} - \frac{1}{2}\lambda_8 v^2 s^2_{\beta} - \frac{1}{4}\lambda_6 v_S^2 \label{eq:ewsb3}
\end{eqnarray}
where $s_x, c_x, t_x$ stand for $\sin x$, $\cos x$ and $\tan x$ respectively, and $\lambda_{345} = \lambda_3 + \lambda_4 + \lambda_5$ and $t_\beta = v_2/v_1$.

After the EWSB of $SU(2)_L\otimes U(1)_Y$ down to electromagnetic $U(1)$, three of the nine Higgs
degrees of freedom corresponding to the Goldstone bosons are absorbed by the longitudinal
components of vector boson $W^\pm$ and $Z^0$. The remaining six degrees of freedom should manifest as physical Higgses: three CP-even scalars ($h_1$, $h_2$, $h_3$ with $m_{h_1}< m_{h_2}< m_{h_3}$), one CP-odd $A$ and a charged Higgs pair $H^\pm$.

\subsection{Higgs masses and mixing angles}
\label{sec:higgsmasses}
The most general form of the squared mass matrix $7\times 7$  of the Higgs sector  
can be recast, using Eqs.(\ref{eq:ewsb1}, \ref{eq:ewsb2}, \ref{eq:ewsb3}), 
into a block diagonal form of three submatrices: one $3 \times 3$ matrices denoted in the following by 
${\mathcal{M}}_{{\mathcal{CP}}_{even}}^2$ for CP-even sector, 
one $2\times 2$ matrix ${\mathcal{M}}_{{\mathcal{CP}}_{odd}}^2$ for CP-odd sector and one $2 \times 2$ matrix denoted by ${\mathcal{M}}_{\pm}^2$ for the charged sector.

The squared mass matrix for the charged fields $\phi_{1,2}^{\pm}$ is:
\begin{eqnarray}
{\mathcal{M}}_{\pm}^2= \left(
  \begin{array}{cc}
  \mu^2 t_\beta - \frac{1}{2}\lambda_{45}^+v^2 s^2_{\beta}\,\,\,&\,\,\,-\mu^2 + \frac{1}{2}\lambda_{45}^+v^2 s_\beta c_\beta\\
  -\mu^2 + \frac{1}{2}\lambda_{45}^+v^2 s_\beta c_\beta\,\,\,&\,\,\,\mu^2 t_\beta^{-1} - \frac{1}{2}\lambda_{45}^+v^2 c^2_{\beta}\\
  \end{array}
\right)
\end{eqnarray}
with $\lambda_{45}^+ = \lambda_{4}+\lambda_{5}$.  This matrix is diagonalized by the following orthogonal matrix
$\mathcal{R}_{\beta}$, given by :
\begin{eqnarray}
{\mathcal{R}}_{\beta} &=& \left(
\begin{array}{cc}
c_\beta & -s_\beta \\
s_\beta & c_\beta \\
\end{array}
\right)
\label{eq:rotamatbetaprime}
\end{eqnarray}
Among the two eigenvalues of ${\mathcal{M}}_{\pm}^2$, one is zero and 
corresponds to the charged Goldstone boson $G^\pm$ while the other one 
corresponds to the charged Higgs boson $H^\pm$ and is given by:
\begin{eqnarray}
m_{H^\pm}^2 & = & \frac{1}{s_{2\beta}}\big[ 2\mu^2 - \frac{1}{2}\lambda_{45}^+v^2 s_{2\beta} \big]
\label{eq:mHpm}
\end{eqnarray}

The  charged Higgs $H^\pm$ and the charged goldstone $G^\pm$ are orthogonal rotation of the weak eigenstates $\phi_1^{\pm}$, $\phi_2^{\pm}$,
\begin{eqnarray}
  G^\pm &=&  c_\beta \phi_1^{\pm}+  s_\beta \phi_2^{\pm} \quad , \quad 
  H^\pm = - s_\beta \phi_1^{\pm}+  c_\beta \phi_2^{\pm}
\end{eqnarray}
%
%
The neutral scalar and pseudo-scalar mass matrices are given by:
\begin{eqnarray}
{\mathcal{M}}_{{\mathcal{CP}}_{even}}^2\hspace{-0.15cm}=\hspace{-0.15cm}
\left(
\begin{array}{ccc}
\mu^2 t_\beta + \lambda_1v^2 c^2_{\beta} & -\mu^2 + \lambda_{345} v^2 s_\beta c_\beta & \frac{\lambda_7vv_S c_\beta}{2\sqrt{2}} \\
-\mu^2 + \lambda_{345} v^2 s_\beta c_\beta & \mu^2 t_\beta^{-1} + \lambda_2 v^2 s^2_{\beta} & \frac{\lambda_8vv_S s_\beta}{2\sqrt{2}} \\
\frac{\lambda_7vv_S c_\beta}{2\sqrt{2}} & \frac{\lambda_8vv_S s_\beta}{2\sqrt{2}} & \frac{\lambda_6v_S^2}{8}
\end{array}
\right)\nonumber\\
\label{matrice_CPeven}
\end{eqnarray}
and
\begin{eqnarray}
{\mathcal{M}}_{{\mathcal{CP}}_{odd}}^2=
\left(
\begin{array}{cc}
\mu^2 t_\beta - \lambda_5v^2 s^2_{\beta}\,   &  \, -\mu^2 + \lambda_{5} v^2 s_\beta c_\beta \, \\
-\mu^2 + \lambda_{5} v^2 s_\beta c_\beta \,  &  \, \mu^2 t_\beta^{-1} - \lambda_5 v^2 c^2_{\beta}
\end{array}
\right)
\label{matrice_CPodd}
\end{eqnarray}
The physical states $h_i = \{h_1,\  h_2,\  h_3\}$
 are obtained by an orthogonal transformation $h_i = \mathcal{R}_{\alpha_{1,2,3}} \phi_i$, $(i=1,2,s)$ that diagonalizes the mass matrix ${\mathcal{M}}_{{\mathcal{CP}}_{even}}^2$,

\begin{eqnarray}
\left(
\begin{array}{ccc}
h_1\\
h_2\\
h_3
\end{array}
\right) = \mathcal{R}_{\alpha_{1,2,3}}\,\left(
\begin{array}{ccc}
\phi_1\\
\phi_2\\
\phi_s
\end{array}
\right)
\end{eqnarray}

with :
\begin{eqnarray}
\mathcal{R}_{\alpha_{1,2,3}} = 
\left(
\begin{array}{ccc}
 c_1 c_2 & s_1 c_2  & s_2 \\
-c_1 s_2 s_3 - s_1 c_3 & c_1 c_3 - s_1 s_2 s_3 & c_2 s_3 \\
 -c_1 s_2 c_3 + s_1 s_3  & -c_1 s_3 - s_1 s_2 c_3 & c_2 c_3
\end{array}
\right)
\label{mrotation}
\end{eqnarray}
with $s_i=\sin \alpha_i \ \ c_i=\cos\alpha_i$. 

\noindent
Without loss of generality we assume that $m_{h_1}< m_{h_2}<m_{h_3}$.

From Eq.(\ref{matrice_CPodd}), it is easy to get the two eigenvalues of ${\mathcal{M}}_{{\mathcal{CP}}_{odd}}^2$, one is vanishing and corresponds to the neutral Goldstone boson $G^0$ while the other one corresponds 
to the pseudo-scalar $A$:
\begin{eqnarray}
m_{A}^2 & = & \frac{1}{ s_\beta c_\beta}\big[ \mu^2 - \lambda_{5}v^2 s_\beta c_\beta \big]
\label{eq:mGA}
\end{eqnarray}

The  CP-odd state $A$ and the neutral Goldstone $G^0$ are obtained by an 
orthogonal rotation of the weak eigenstates $\chi_1$, $\chi_2$:
\begin{eqnarray}
G^0 &=&  c_\beta \chi_1+  s_\beta \chi_2\quad , \quad
A= - s_\beta \chi_1+  c_\beta \chi_2
\end{eqnarray}

\subsection{Yukawa texture}
\label{sec:yukawatexture}
There are different types of Higgs couplings to fermions. If we do like in the SM and allow both Higgs fields to couple to all fermions through the following lagrangian:
\begin{widetext}
\begin{eqnarray}
\mathcal{L}_Y&=&\overline{Q}^0_L\tilde{\Phi}_2\eta^{U,0}_2U^0_R+\overline{Q}^0_L{\Phi}_2\eta^{D,0}_2D^0_R+\overline{Q}^0_L\tilde{\Phi}_1\eta^{U,0}_1U^0_R+\overline{Q}^0_L{\Phi}_1\eta^{D,0}_1D^0_R
+ {\rm h.c} 
\end{eqnarray}
\end{widetext}
where $Q_L^0$ is the weak isospin quark doublet, $U_R^0$ and $D_R^0$ are the weak isospin quark singlets 
and $\eta_{{1,2}}^{U,0}$, $\eta_{{1,2}}^{D,0}$ are matrices in flavor space, then the above lagrangian will generate Flavor Changing Neutral Currents (FCNC) at the tree level which 
can invalidate some low energy observables in B, D and K physics.
In order to avoid such FCNC, it is customary to invoke a $Z_2$ symmetry that forbids FCNC couplings at
the tree level \cite{weinberg}.  Depending on the $Z_2$ assignment, we have four type of models \cite{Branco}. In the present study, we focus only on type-I where all fermions couple only to one of the two Higgs doublets. In this case:
\begin{widetext}
\begin{eqnarray}
\mathcal{L}_Y^I=\overline{Q}^0_L\tilde{\Phi}_2\eta^{U,0}_2U^0_R+\overline{Q}^0_L{\Phi}_2\eta^{D,0}_2D^0_R +\overline{U}^0_R
\eta^{U,0\dagger}_2\tilde{\Phi}_2^\dagger Q^0_L+\overline{D}^0_R \eta^{D,0\dagger}_2 \Phi_2^+ Q^0_L
\end{eqnarray}
\end{widetext}

Neutral Higgs couplings to a pair of fermions are:
\begin{eqnarray}
h_1 f\overline{f} & & \frac{\mathcal{R}_{12}}{s_\beta} = \frac{c_2 s_1}{s_\beta} \nonumber \\
h_2 f\overline{f} & & \frac{   \mathcal{R}_{22}}{s_\beta} = \frac{(c_1c_3 -s_1s_2s_3)}{s_\beta} \nonumber \\
h_3 f\overline{f} & & \frac{   \mathcal{R}_{32}}{s_\beta} = -\frac{(c_1s_3 +s_1s_2c_3)}{s_\beta}
\label{table1}
\end{eqnarray}
where $f$ designate any type of fermions.

\subsection{Higgs couplings to gauge bosons and sum rules}
We present shortly here the Higgs couplings to gauge bosons and discuss the sum rules  required by unitarity
\cite{Gunion:1990kf,Bento:2017eti} that are subject to. 
The normalized couplings of neutral Higgs to a pair of gauge bosons $V=Z, W$ are given by:

\begin{alignat}{2}
&g_{h_1 V V}: &\quad &  c_\beta  \mathcal{R}_{11}+s_\beta  \mathcal{R}_{12}   = c_{\alpha_2} c_{\beta-\alpha_1},\label{align1}\\
&g_{h_2 V V}: & \quad &  c_\beta  \mathcal{R}_{21}+s_\beta  \mathcal{R}_{22} 
= c_{\alpha_3} s_{\beta-\alpha_1} - s_{\alpha_2} s_{\alpha_3} c_{\beta-\alpha_1},\label{align2}\\
&g_{h_3 V V}: & \quad & c_\beta  \mathcal{R}_{31}+s_\beta  \mathcal{R}_{32} 
=- s_{\alpha_3} s_{\beta-\alpha_1}- s_{\alpha_2} c_{\alpha_3} c_{\beta-\alpha_1},\label{align3}
\end{alignat}
which satisfy the following sum rule:
\begin{eqnarray}
\sum_{i=1}^3 g_{h_i V V}^2 =1
\label{sumrulehvv}
\end{eqnarray}
This sum rule imply that each coupling $g_{h_iVV}$ is requested to satisfy: $\mid g_{h_iVV}\mid \leq 1$.

\noindent
For the couplings between two Higgs bosons and one gauge boson, we can distinguish two cases, a neutral case which corresponds to $h_i A Z $ vertex and charged case associated with  $h_i H^\mp W^\pm$ vertex. 
From the kinetic terms of the Higgs fields, one can derive the various trilinear couplings among neutral, charged Higgses and gauge bosons. 
In units  of $\lambda_{n}=\frac{\sqrt{g^2+{g^{'}}^2}}{2}(p_{h_i}-p_{A})_\mu$ for neutral Higgs, 
respectively in units of  $\lambda_{c}=\mp\frac{g}{2}(p_{h_i}-p_{H^\pm})_\mu$ for charged Higgs, we have:
\begin{alignat}{2}
&g_{h_1 V S}: & \quad & -c_{\alpha_2} s_{\beta-\alpha_1},\label{align4}\\
&g_{h_2 V S}: & \quad & c_{\alpha_3} c_{\beta-\alpha_1} + s_{\alpha_2} s_{\alpha_3} s_{\beta-\alpha_1},\label{align5}\\
&g_{h_3 V S}: & \quad & -s_{\alpha_3} c_{\beta-\alpha_1} + s_{\alpha_2} c_{\alpha_3} s_{\beta-\alpha_1},\label{align6}
\end{alignat}
where V=Z and $S=A$ for the neutral case and $V=W^\pm$ and $S=H^\mp$ for charged one.

In the 2HDMS, one can easily derive the following sum rules:
\begin{eqnarray}
&&  g_{h_i W^\pm W^\mp}^2 + g_{h_i W^\pm H^\mp}^2 + R_{i3}^2=1 , \quad  i=1,2,3 \label{align7}\\
&&  g_{h_i Z Z}^2 + g_{h_i Z A}^2 + R_{i3}^2=1 \quad , \quad i=1,2,3 \label{align8}
\label{sumrule}
\end{eqnarray}
Where $R_{i3}$ is the singlet component of the Higgs $h_i$. 

An other type of sum rule which relate $h_if\overline{f}$ and $h_iVV$ can be derived from the Feynman rules and is given by \cite{Bento:2018fmy}:
  \begin{eqnarray}
&& \sum_{i=1}^3  g_{h_iVV} g_{h_if\overline{f}}=1
\label{sumrulevf}
\end{eqnarray}
where $g_{h_iVV}$ and $g_{h_if\overline{f}}$
are the normalized couplings of $h_i$ to gauge bosons and fermions.

From above, it follows that:
\begin{itemize}
\item  if $h_i$ is pure singlet ($R_{i3}^2\approx 1$), then from eqs.(\ref{sumrule}) 
one has $g_{h_i WW}^2 + g_{h_i W^\pm H^\mp}^2\approx 0$  and
$g_{h_i ZZ}^2 + g_{h_i ZA}^2\approx 0$ which would imply that 
$h_iVV$, $h_iH^\pm W^\mp$ and $h_iAZ$ must be very suppressed, 
and this will present a real challenge for the production and detection of such Higgs bosons.

\item if $g_{h_iVV}=1$ which means  that $h_i VV$ is full strength, then both singlet component $R_{i3}$ as well as 
$g_{h_i SV}$ couplings must vanish. This scenario could happen only when $h_i$ have no singlet component.

\item According to Eq.(\ref{sumrulehvv}), if $g_{h_iVV}=1$ then $g_{h_jVV}=0$ for $j\neq i$. This would imply from 
Eq.(\ref{sumrulevf}) that the reduced coupling to fermions must satisfy $g_{h_if\overline{f}}=1$.
\end{itemize}

\section{Theoretical and experimental constraints}
\label{constraints}
The Two Higgs Doublets Model plus a Singlet possesses a large freedom in the scalar sector, 
coming from the large number of free parameters of the scalar potential. 
In order to obtain a viable model, many theoretical constraints have to be imposed on the scalar
potential like perturbative unitarity, vaccum stability and electroweak precision observables. 
In what follows, we will describe briefly these constraints.

\subsection{Boundedness from below (BFB) of the potential}
\label{bfb}
In order to ensure a stable vacuum, the scalar potential has to be bounded from below
in any directions in the field space as the field strength becomes extremely large.
At large field values, the scalar potential is fully dominated by quartic couplings 
whose the BFB will depend only.

At large field strength, the potential defined by Eq.(\ref{eq:Vpot}) is generically dominated by the quartic terms: 
\begin{widetext}
\begin{eqnarray}
V^{(4)}(H_1,H_2,S) &=& \frac{\lambda_1}{2} \left( H_1^\dagger H_1 \right)^2
+ \frac{\lambda_2}{2} \left( H_2^\dagger H_2 \right)^2 + \lambda_3\, H_1^\dagger H_1\, H_2^\dagger H_2 + \lambda_4\, H_1^\dagger H_2\, H_2^\dagger H_1\nonumber\\
&+&\frac{\lambda_5}{2} \left[\left( H_1^\dagger H_2 \right)^2+ \left( H_2^\dagger H_1 \right)^2 \right] \nonumber\\
&+&\frac{1}{8} \lambda_6 S^4 +\frac{1}{2} \lambda_7 \left( H_1^\dagger H_1 \right) S^2+\frac{1}{2}\lambda_8 \left( H_2^\dagger H_2 \right) S^2
\label{eq:Vquartic}
\end{eqnarray}
\end{widetext}
\noindent 
The study of $V^{(4)}(H_1,H_2,S) $ will thus be sufficient to obtain the main constraints. 
The full BFB constraints reads as 
\begin{eqnarray}
&&\lambda_1\,,\,\lambda_2\,,\,\lambda_6 > 0 \quad , \quad \lambda_3 + \sqrt{\lambda_1\lambda_2} > 0 \label{new2hdmS_2}  \\
&&\lambda_3 + \lambda_4 - |\lambda_5| + \sqrt{\lambda_1\lambda_2} > 0 \label{new2hdmS_3} \\
&&  \lambda_7 > -\sqrt{\lambda_1\lambda_6} \quad , \quad  \lambda_8 > -\sqrt{\lambda_2\lambda_6} 
\end{eqnarray}
for $\lambda_7 >0 \, \text{and}\, \lambda_8 > 0 $.\\ 
 
If $\lambda_7\, \text{or}\, \lambda_8 < 0$, we have to satisfy two additional constraints: 
\begin{eqnarray}
&&\hspace{-0.2cm}\lambda_3\,\lambda_6 - \lambda_7\lambda_8 + \sqrt{(\lambda_1\lambda_6 - \lambda_7^2)(\lambda_2\lambda_6 - \lambda_8^2)} > 0   \\
&&\hspace{-0.2cm}\lambda_6\,(\lambda_3+\lambda_4 +|\lambda_5|) - \lambda_7\lambda_8 + \sqrt{(\lambda_1\lambda_6 - \lambda_7^2)(\lambda_2\lambda_6 - \lambda_8^2)} > 0\nonumber\\
\end{eqnarray}
Full technical details on the proof of  these constraints can be found in Appendix.(\ref{appendix-bfb}).

\subsection{Perturbative unitarity}
\label{unitarity}
To constrain further the scalar potential parameters of the 2HDMS one can ask that 
tree-level perturbative unitarity is preserved for a variety of scattering processes: 
gauge boson-gauge boson scattering
scalar-scalar scattering, and also scalar-gauge boson scattering. 
Moreover, according to the equivalence theorem which states that at high energy limit $\sqrt{s}$ the amplitudes of a
scattering process involving longitudinally polarized gauge bosons V  are asymptotically equal, up to correction 
of the order $m_V/\sqrt{s}$, to the corresponding scalar amplitudes in which   longitudinally polarized  
gauge bosons are replaced by their corresponding Goldstone bosons. 
We conclude that perturbative unitarity constraints can be implemented by considering 
pure scalar-scalar scattering only.

In order to derive the perturbative unitarity constraints on the scalar parameters of 2HDMS we follow 
refs \cite{Kanemura:1993hm,arhrib}.
According to \cite{Kanemura:1993hm,arhrib}, one computes the scattering amplitude in the weak eigenstate basis where the quartic couplings are less involving (does not involve mixing angles $\alpha_i$ and $\beta$). 
The important point is that the amplitude expressed in the mass eigenstate fields can be transformed into the amplitude for the non-physical fields by making a 
unitary transformation. The eigenvalues for the scattering amplitude should be unchanged under such a unitary transformation.

In the Appendix.(\ref{appendix-unitarity}) we present the technical details of the different 
$2\to 2$ scattering amplitudes. 
The explicit forms of the eigenvalues at tree level are given by:

\begin{eqnarray}
|\lambda_3+\lambda_4|\ ,\ |\lambda_3\pm\lambda_5|\ , \ |\lambda_3+2\lambda_4 \pm 3\lambda_5| < 8\pi \nonumber \\
\big|\frac{\lambda_7}{2}\big|\ , \ \big|\frac{\lambda_8}{2}\big|\ , \ |\frac{\lambda_6}{4}|< 8\pi \nonumber \\
\big|\frac{1}{2}(\lambda_1+\lambda_2\pm\sqrt{(\lambda_1-\lambda_2)^2+4\lambda_4^2})\big|< 8\pi \nonumber \\
\big|\frac{1}{2}(\lambda_1+\lambda_2\pm\sqrt{(\lambda_1-\lambda_2)^2+4\lambda_5^2})\big|< 8\pi \nonumber
\end{eqnarray}
Other eigenvalues are coming from the cubic polynomial equation associated to the submatrix 
${\cal M}_2$ corresponds to scattering with one of the following
initial and final states:
$(\phi_1^+\phi_1^-$, $\phi_2^+\phi_2^-$, $\frac{\phi_1\phi_1}{\sqrt{2}}$,$\frac{\phi_2\phi_2}
{\sqrt{2}}$,$\frac{\phi_s\phi_s}{\sqrt{2}}$,$\frac{\chi_1\chi_1}{\sqrt{2}}$,$\frac{\chi_2\chi_2}{\sqrt{2}})$. 
For more details, see Appendix.(\ref{appendix-unitarity}),

Moreover, we also force the potential to be perturbative by imposing that all quartic couplings of the scalar potential satisfy $|\lambda_i| \leq 8 \pi$ ($i=1,...,8$).

\subsection{Electroweak precision test observables (EWPT)}
The oblique parameters S, T, and U are known to provide an indirect probe of new physics BSM for theories that process  $SU(2) \times U(1)$ symmetry~\cite{Peskin}.  These parameters
quantify deviations from the SM in terms of radiative corrections to the W, Z and the photon self-energies. 
In the framework of 2HDMS, the Higgs doublet couples to the W and Z gauge bosons via the covariant derivative. 
Due to singlet and doublet admixtures in the scalar sector, the singlet field will also couple to the gauge bosons W and Z. 
Therefore, both neutral Higgs $h_i$, $A$ and charged Higgs will contribute to S and T 
parameters which are very well constrained by electroweak precision test observables. These EWPT constraints will be converted to limit on the mixing angles and/or masses splitting among the 2HDMS spectrum. The extra-contribution to S, T and U parameters for 2HDMS are given in Appendix. (\ref{Oblic}). 

In order to study the correlation between S and T, we perform the $\chi^2$ test over the allowed parameter space of 2HDMS. Our $\chi^2_{S, T}$ is defined as:

\begin{widetext}
\begin{equation}
\chi^2_{S,T} = \frac{1}{\hat{\sigma}^2_{1}(1-\rho^2)}(S - \hat{S})^2
+ \frac{1}{\hat{\sigma}^2_{2}(1-\rho^2)}(T - \hat{T})^2 - \frac{2\rho}{\hat{\sigma}_{1}\hat{\sigma}_{2}(1-\rho^2)}(S - \hat{S})(T - \hat{T}) \,, \label{eq:chi2}
\end{equation} 
\end{widetext}
where S and T are the computed quantities within 2HDMS framework~\cite{Grimus,Lavoura,G.2003}. $\hat{S}$ and  $\hat{T}$ are the measured values of S and T, $\hat{\sigma}_{1,2}$  are their one-sigma errors and $\rho$  their correlation~\cite{Haller 2018}, 

\begin{equation}
S = 0.04 \pm 0.11, \quad T = 0.09\pm 0.14, \quad \rho_{S,T} = 0.92
\end{equation}
It is worth noting here that we have checked the limits on the oblique parameters with the THDM, in this sense, our results match exactly to those outlined in \cite{Su:2001,Kanemura.2011}.

In addition, we have indirect experimental constraints 
from $B$ physics observables on the contribution of the 2HDMS such as $\tan\beta$ and $m_{H^\pm}$.
In the 2HDMS, the charged Higgs coupling to fermions is not at all affected by the singlet component of the additional Higgs.
Therefore, constraints from $B\to X_s\gamma$ and $B_q$ mixing will be the same as for the usual 2HDM model.
We remind the reader that the recent experimental results presented by the Heavy Flavor Averaging Group (HFAG)~\cite{Amhis} of $B(B\rightarrow X_s \gamma)$ have changed in a significant way the bounds on the charged Higgs boson mass.
For instance, in the Type-II 2HDMS, the measurement of the BR$(B \to X_s\gamma)$ constrains charged Higgs mass 
to be larger than about 570 GeV \cite{misiak}, while in Type-I 2HDMS one can still obtain a charged Higgs boson with a 
mass as low as  $100-200$ GeV  provided that $\tan \beta \geq 2$.

\subsection{Constraints from Higgs data}
Both ATLAS and CMS experiments of the LHC Run1 with 7 and 8 TeV and Run2 with 13 TeV 
confirmed the discovery of a Higgs-like particle with a mass around 125 GeV.
Both groups performed several measurements on the Higgs-like particle couplings to the SM
particles. Recently, both ATLAS and CMS Collaborations have announced the observation of Higgs 
bosons produced together with a top-quark pair \cite{Sirunyan:2018hoz,Aaboud:2018urx}. 
All these measurements seem to be in perfect agreement with SM predictions.

In the case of 2HDMS, all tree level Higgs couplings to fermions and gauge bosons are 
modified with the mixing parameters $\alpha_i$ and $\beta$. 
The loop mediated processes such as $gg\to h_i$, $h_i\to \gamma\gamma$ and $h_i\to \gamma Z$
 will be affected both by the mixing angles as well as by the  additional charged 
 Higgs $H^\pm$ loops  which depend on the triple scalar coupling $h_iH^\pm H^\mp$.

To study the effects of ATLAS and CMS  measurements on 2HDMS, 
we take into account experimental data from the observed cross section times branching ratio divided by the corresponding SM predictions for the various channels, i.e. the signal strengths of the Higgs boson defined by:

\begin{equation}
\mu_i^f = \frac{\sigma_i(h) Br(h \rightarrow f)}{\sigma_i^{SM}(h) Br^{SM}(h \rightarrow f)}
\end{equation}
where $i$ stand for different modes of Higgs production. The dominant mechanisms of Higgs production are gluon fusion (ggF), followed by vector boson fusion (VBF), Higgs-strahlung (Vh) and associated production with top-quark pairs ($\overline{t}th$). 
 
All these various signal strength channels are included in our analysis through the public code 
HiggsBounds and HiggsSignals which also include previous LEP and Tevatron experimental searches.

As said previously, in our analysis, we will assume that $h_2$ is the 125 GeV Higgs-like particle discovered while $h_1$ would be lighter than $h_2$. Therefore, once  the decay channels $h_2\to h_1h_1$ and/or $h_2\to AA$ are open, the subsequent decays of $h_1/A$  into fermions, photons or gluons, will lead either to invisible or undetected $h_2$ decays that can be constrained by using global analysis to the present ATLAS and CMS data to Higgs couplings.

We stress here that there is also searches for non-detected decays of the SM Higgs boson both by ATLAS and CMS.
CMS look for the following SM Higgs production channels: gluon fusion, vector boson fusion, and Higgsstrahlung process $pp\to VH$ (V=W or Z) with subsequent invisible Higgs decays. Upper limits are placed on 
$Br(H\to invisible)$, as a function of the assumed production cross sections. 
The combination of all the above channels, assuming SM production, 
yields to an upper limit of 0.24 on the  $BR(H\to invisible)$ at the 95$\%$ confidence level 
\cite{Khachatryan:2016whc}. ATLAS collaboration performs a search for an invisible decay of the Higgs 
through $pp\to ZH$ process  with a leptonic subsequent decay of the Z \cite{Aaboud:2017bja}. 
Their limit is slightly weaker than CMS results.

In our study, we will use the fact that the total branching fraction of the SM-like Higgs boson into undetected BSM decay modes is constrained, as mentioned, by $BR(H\to invisible) \leq 0.24$ where 
$BR(H\to invisible)$ designate $BR(h_2 \to h_1h_1)$ or the sum of $BR(h_2 \to h_1h_1)$ 
and $BR(h_2 \to AA)$,  $BR(h_2 \to Z^*A)$ and $BR(h_2 \to W^* H^\pm)$ if the later is open.

\section{Numerical results}
\subsection{Parameters scan}
The scalar potential Eq.(\ref{eq:Vpot}) have 15 independent parameters: four masses, 
8 quartic couplings $\lambda_{1,...,8}$ and 3 vaccum expectation values. Three masses can 
be eliminated by the use of the 3 minimization conditions Eq.(\ref{eq:ewsb3}).
Moreover, after electroweak symmetry breaking, from the kinetic terms of the Higgs doublets, 
the W and Z gauge bosons acquire masses which are given by $m_W^2 = \frac{1}{2}g^2(v_1^2+v_2^2)$ and 
$m_Z^2 = \frac{1}{2}(g^2 +g'^2) (v_1^2+v_2^2)$, where $g$ and $g'$ are the $SU_L(2)$ and $U_Y(1)$ gauge
couplings. The combination $v_1^2 +v_2^2$ is thus fixed by the electroweak scale
through the well known relation $v^2=v_1^2 + v_2^2 =(2\sqrt{2}G_F)^{-1}$,  
and we are left with 11 free parameters.
By simple algebraic calculations, from the mass matrix relations, 
one can express all the quartic couplings $\lambda_i$ as a function of the physical masses, 
$\mu^2$, $\tan\beta$ and the mixing angles $\alpha_i$ \footnote{In Ref.~\cite{rui2017} one can find the expressions of all $\lambda_i$s as well as the trilinear and quartic scalar couplings as a function 
 of the physical masses and the mixing angles.}. 
 One can then take the following set of free independent parameters: 
\begin{center}
$\alpha_{1,2,3}$,\, $\tan\beta$,\, $v_S$,\, $m_{h_{1,2,3}}$,\, $m_{A}$,\, 
$m_{H^\pm}$\, \ \ {\rm and}\, $\mu^2$
\end{center}
with the convention $m_{h_1}  < m_{h_2}  < m_{h_3}$. 

Note that the usual 2HDM is recovered from 2HDMS by taking the following limits:
\begin{eqnarray}
&& \alpha_1 \to \alpha+\frac{\pi}{2} \ \ , \ \ \alpha_2 \to 0 \ \  and \ \ \alpha_3 \to 0
\label{2hdmlimit} 
\end{eqnarray}

In the previous studies on the 2HDMS \cite{rui2017,Engeln:2018mbg}, 
they concentrate on $h_1$ being the 125 GeV SM-like Higgs
while in our analysis we will study the consistency of having the second Higgs $h_2$ as the 125 GeV SM-like Higgs.
This scenario open more decay channels for $h_2$ such as $h_2\to h_1 h_1$ and probably 
$h_2\to AA, AA^*, Z^*A, W^*H^\pm$.

In order to display the allowed regions for the parameters, we have considered both of the exclusions from both HiggsBounds-5.1beta and HiggsSignals-2.1.0beta to compute the value of $\chi^2_{min}$ considering the combination of 8 TeV and 13 TeV Higgs signal strength data from run-I and run-II. Thus, we show the best fit at 
$95.5\%$ C.L, which corresponds to $2.3\leq\Delta\chi^2 (\chi^2-\chi^2_{min}) \leq 5.99$.

\begin{figure*}[!h]
\centering
\resizebox{0.32\textwidth}{!}{
\includegraphics{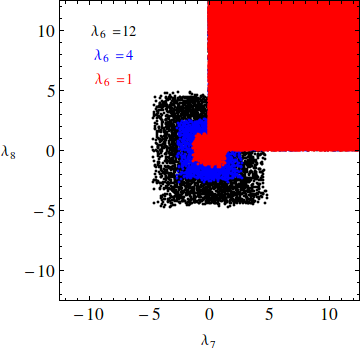}}
\resizebox{0.32\textwidth}{!}{
\includegraphics{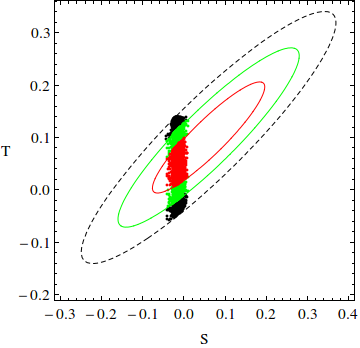}}
\caption{The left panel shows the parameter space allowed by the theoretical 
constraints(Uniatrity, Perturbativity and BFB) for ($\lambda_7,\lambda_8$) for $\lambda_6 =(1,4,12)$. 
Note the complete overlap between black, blue and red for positive $\lambda_{6,7}$.
The right panel illustrates the correlation between oblique parameters S and T. 
The errors for $\chi^{ST}$-square fit are 99.7$\%$ CL (black),  95.5$\%$ CL (green) and 68$\%$ CL (red)}
\label{fig:constraintes}
\end{figure*}

In the left panel of Figure.(\ref{fig:constraintes}), we show the allowed parameter for 
$\lambda_{7,8}$ for various values of $\lambda_6$ taking into account perturbative unitarity and BFB constraints. 
We note first that there is a complete overlap between the three colors for positive $\lambda_{7,8}$.
One can see that for negative $\lambda_{7,8}$,  the theoretical constraints restrict their strength for small $\lambda_6$.  The restriction is relaxed for large $\lambda_6\approx 4\pi$.   
 
In the right panel of Figure.(\ref{fig:constraintes}), we present the correlation between S and T, 
after taking into account the theoretical constraints and the exclusion from Higgs Bounds at 95\% C.L. The red, green and black points represent the points with $\Delta\chi^2$ value within the 
$1 \sigma$, $2\sigma$ and $3\sigma$ interval.

\subsection{Results for 2HDMS type I}
In this section, we study the case of 2HDMS type I, where both $h_1$ and $A$ could be rather light. 
We scan over the following range:

\begin{widetext}
\begin{eqnarray}
&& m_{h_1}\in [10, 120]\ \text{GeV}, \ \ m_{h_3}\in [200, 700]\ \text{GeV},  \ \ m_{H^\pm}\in [80, 700]\ \text{GeV}, \nonumber \\
&& m_{A}\in [62.5, 700]\ \text{GeV}, \ \ \mu^2\in [0, 1.5\times10^3]\ \text{GeV}, \ \
v_{S} = 300\ \text{GeV}, \nonumber \\
&& \frac{-\pi}{2}<\alpha_{1}<\frac{\pi}{2},  \ \ \frac{-\pi}{6}<\alpha_{2,3}<\frac{\pi}{6}, \ \ and  \ 0.5<\tan\beta<25, 
\label{param-typeI}
\end{eqnarray}
\end{widetext}

In our scan we allow $h_1$ to be in the range $[10,124]$ GeV 
while $m_A\geq 62.5$ GeV. 
In such configuration, only $h_2\to h_1h_1$ and/or $h_2\to AZ^*,H^\pm W^{\mp *}$ could be open. $h_2\to AA^*\to Aff'$ is suppressed both by the phase space and also by the coupling of $A$ to light fermions.  
Since $h_2$ is identified as the SM-Higgs-like, the non detected decays such as $h_2\to h_1h_1, AA^*, AZ^*,H^\pm W^{\mp *}$ if open should not exceed 24\% as we explain below.

In order to be consistent with the EW precision measurements, such light $h_1$ and A are naturally 
accompanied also by a light charged Higgs which is consistent with $B\to X_s\gamma$ constraint in 2HDMS of type I.
\subsection*{$h_1$ decays}
We study first the decay of $h_1$ into SM particles. As one can see from the couplings of $h_1$ to fermions given in Eq.(\ref{table1}), $h_1$ could be fermiophobic if $R_{12}\propto \cos\alpha_2 \sin\alpha_1$ vanish. This scenario
might happens if we take $\alpha_1\approx 0$ and/or $\alpha_2 \approx \pi/2$ which is possible since both $\alpha_1$  and $\alpha_2$ are free parameters in this model.  
\begin{itemize}
\item The case where $\alpha_2=\pi/2$ corresponds to $h_1$ being pure singlet and will not be discussed here. 
\item The case where $\alpha_1=0$ with $\alpha_2\neq \pi/2$, $h_1$ contains both 
doublet and singlet component.
\end{itemize}

\begin{figure*}[!h]
\centering
\resizebox{0.32\textwidth}{!}{
\includegraphics{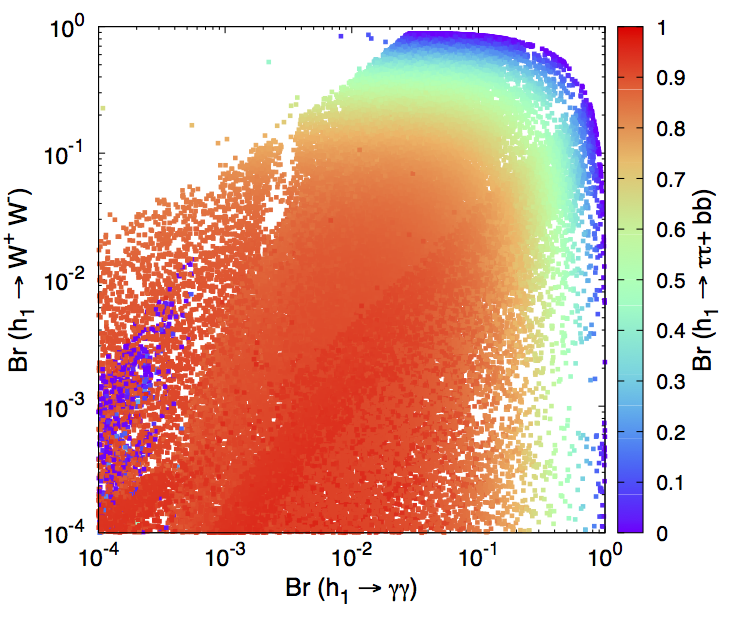}}
\resizebox{0.32\textwidth}{!}{
\includegraphics{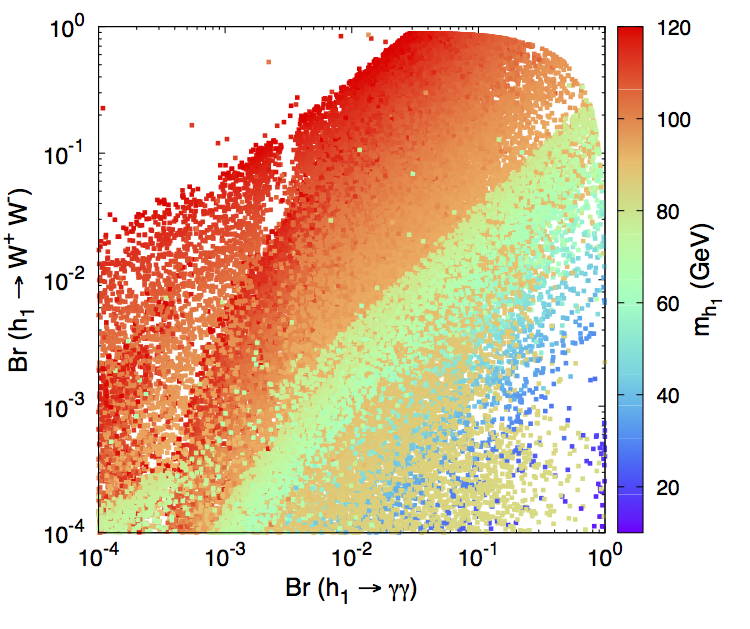}}
\caption{$Br(h_1\rightarrow W^+W^-)$ and $Br(h_1\rightarrow \gamma \gamma)$ 
vs. $Br(h_1\rightarrow b\overline{b}, \tau^+\tau^-)$ (left), and $m_{h_1}$ (right) at 95\% $.CL$ }
\label{fig:type1-gb}
\end{figure*}

In Figure.(\ref{fig:type1-gb}), we show the branching ratio of $h_1\to W^+W^-$ as a function of 
 $h_1\to \gamma\gamma$ with $Br(h_1\to b\overline{b}+\tau^+\tau^-)$ represented on the horizontal axis on the 
 left panel while on the right panel we show $m_{h_1}$. Since   $m_{h_1}\leq 125$ GeV, $h_1\to W^+W^-$ will proceed with one or both W being off-shell.  We first mention that the singlet component of $h_1$ 
 does not exceed 50\% in our case which makes $h_1$ dominated by doublet components.

As one can see, in most of the cases $h_1$ would decay significantly into 
a bottom pair unless $\alpha_1$ vanish which is the fermiophobic limit for $h_1$. 
In this case, it is clear that $h_1\to \gamma\gamma$ reach its maximum value 
when  $Br(h_1\to b\overline{b})$ and  $Br(h_1\to \tau^+\tau^-)$  are very suppressed. 
When $h_1$ is fermiophobic, $h_1\to VV^*$ or $h_1\to V^*V^*$, $V=Z,W$ 
can compete with $h_1 \to \gamma \gamma$. In what follow discuss only $h_1\to WW$ since $h_1\to ZZ$ 
is smaller.  In fact, $h_1\to W^*W^*$ which is open for $m_{h_1}<m_W$  is very suppressed due to  phase space
while for $m_{h_1}\geq m_W$,  $h_1\to WW^*$ is open and could strongly compete with
$h_1 \to \gamma \gamma$.
This is shown in the left and right panel of Figure.(\ref{fig:type1-gb})
where we can see $Br(h_1\to WW^*)$ as a function of $Br(h_1\to \gamma\gamma)$.
Close to the fermiophobic limit
where $h_1\to b\overline{b}$ and $h_1\to \tau^+\tau^-$ are suppressed,  if the mass of $h_1$ is larger than 
the W boson mass then $h_1\to WW^*$ can dominate $h_1\to \gamma \gamma$ in some cases.

\begin{figure*}[!h]
\centering
\resizebox{0.32\textwidth}{!}{
\includegraphics{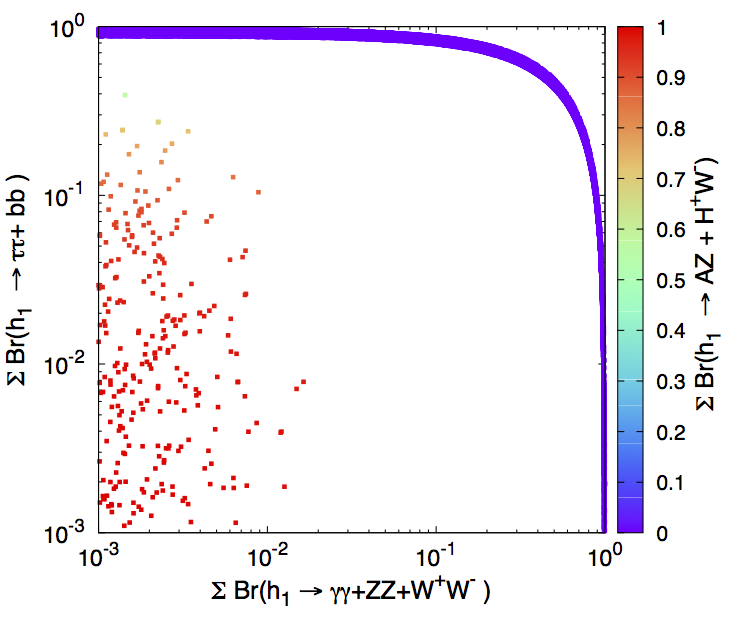}}
\resizebox{0.32\textwidth}{!}{
\includegraphics{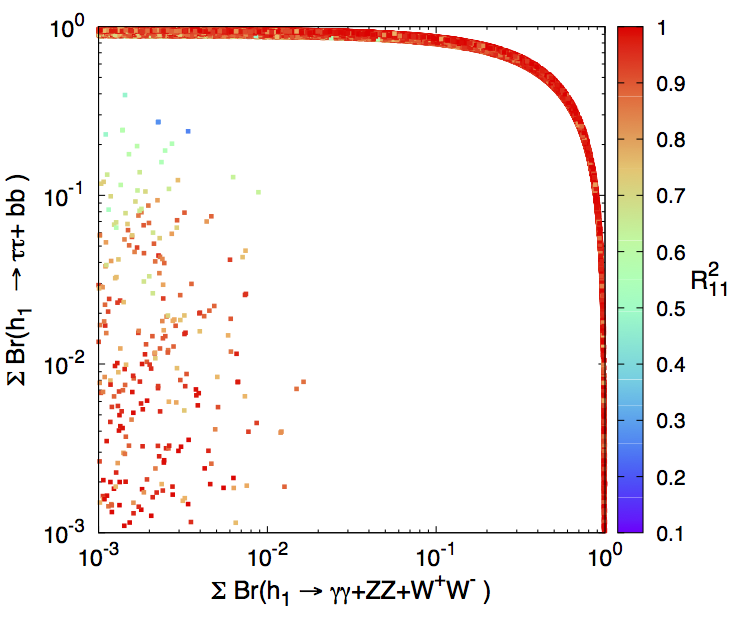}}
\caption{The plot shows $\sum Br(h_1\rightarrow \tau^+ \tau^- + b\overline{b})$ as a function of $\sum Br(h_1\rightarrow \gamma \gamma+ZZ+ W^+W^-)$  vs $Br(h_1 \rightarrow (H^{\pm}W^\mp + AZ)$ (left)  and versus $R_{11}^2$ in the right panel at 95\% $.CL$ in 2HDMS type-I}
\label{ fig:type1-gb2}
\end{figure*}

One could have also the following scenario: both  $h_1\to f\overline{f}$, $h_1\to VV^*$ and $h_1\to \gamma\gamma, \gamma Z$ are rather small while the branching ratio of $Br(h_1\to A Z^*)+Br(h_1\to H^\pm W^{*\mp})$ becomes significant 
which can be understood from the sum rules eqs. (\ref{sumrule}).  Due to the smallness of $h_1VV$ coupling and 
$R_{13}$ component, the sum rule eqs. (\ref{sumrule}) imply that $h_1AZ$ and $h_1W^\pm H^\mp$ could becomes significant.
This is illustrated in the left panel of Figure.(\ref{fig:type1-gb}) with the blue dots in the left-down corner where both 
$Br(h_1\to W W^*) \approx Br(h_1\to \gamma\gamma)\approx 10^{-4}$ and also 
$\sum Br(h_1\rightarrow \tau^+ \tau^- + b\overline{b})$ are rather small.

The above configuration is illustrated clearly in Figure.(\ref{ fig:type1-gb2}) where we show the correlation between
$\sum Br(h_1\rightarrow \tau^+ \tau^- + b\overline{b})$  and 
$\sum Br(h_1\rightarrow \gamma \gamma+ZZ+ W^+W^-)$ 
as a function of $Br(h_1 \rightarrow H^{\pm}W^\mp + AZ)$ on the horizontal axis. 
It can be seen that when the fermionic ($\tau^+\tau^-$, $b\overline{b}$) and bosonic ($\gamma\gamma$, 
$W^+W^-$, ZZ)  decays of $h_1$ are suppressed, the higgs to higgs decays  $h_1 \rightarrow H^{\pm}W^\mp$ and/or 
 $h_1 \rightarrow ZA$ becomes significant. In the right panel of Figure.(\ref{ fig:type1-gb2}) 
 it can be read that $h_1$ is most of the case having large doublet component.

\begin{figure*}[!h]
\centering
\resizebox{0.32\textwidth}{!}{
\includegraphics{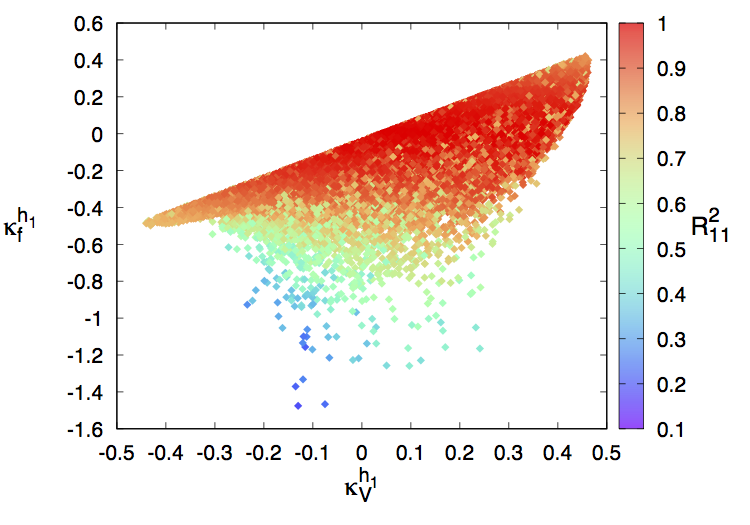}}
\resizebox{0.32\textwidth}{!}{
\includegraphics{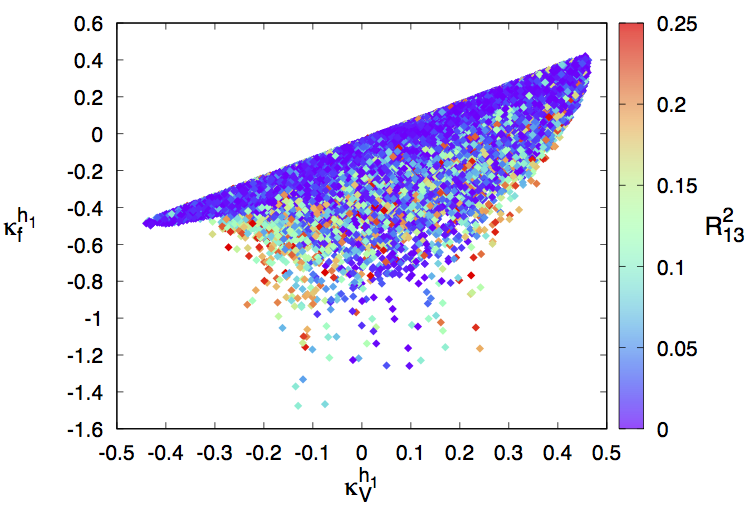}}
\resizebox{0.28\textwidth}{!}{
\includegraphics{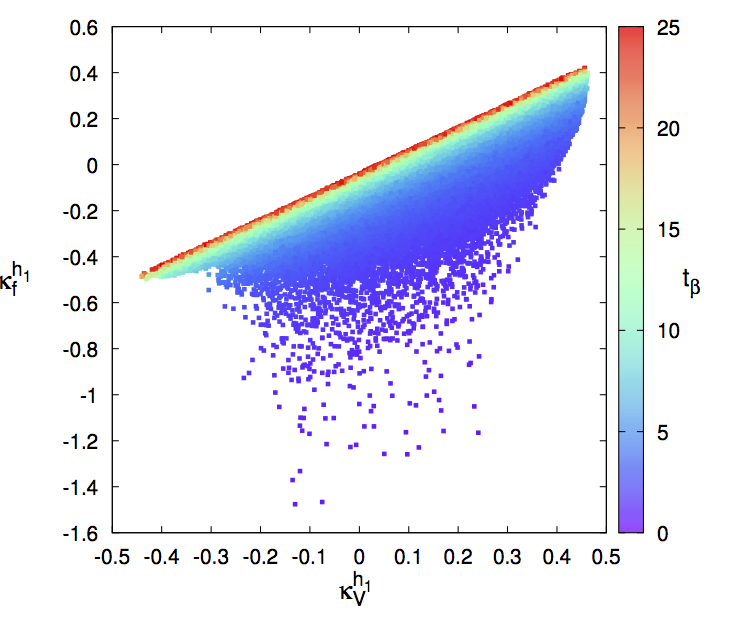}}
\caption{$\kappa_f^{h_1}$ as a function of $\kappa_V^{h_1}$ with $R_{1i}^2$ (i=1,3) on the 
horizontal axis (right and middle panel), and  with $\tan\beta$ on the left panel.}
\label{fig:type1-az-hw}
\end{figure*}

In Figure.(\ref{fig:type1-az-hw}) we illustrate 
$\kappa_f^{h_1}$ as a function of  $\kappa_V^{h_1}$ with $R_{1i} (i=1,2,3)$ component of 
$h_1$ on the horizontal axis.
From this plot, one can read that the doublet component is rather large in most of the cases 
leaving only small singlet component which is less than 50\% . One can also learn 
that when  $\kappa_f^{h_1}$ and  $\kappa_V^{h_1}$ are suppressed, the doublet component is 
very large. Which means that $h_1$ is mainly coming from the doublet components. 
According to the sum rule Eq.(\ref{sumrulehvv}), $\mid \kappa_V^{h_1}\mid \leq 1$ is requested 
which is consistent with Figure.(\ref{fig:type1-az-hw}). 
On the other hand, in large area of parameter space $\mid \kappa_f(h_1)\mid \leq 1$ while one can have 
a scenario with $\mid \kappa_f(h_1)\mid  \geq 1$. This happens when $h_1f\overline{f}\propto R_{12}$ is rather large (small $R_{11}$). On the left panel of Figure.(\ref{fig:type1-az-hw}) we show the sensitivity to 
$\tan\beta$ where we can see a linear correlation between $\kappa_f^{h_1}$ and  $\kappa_V^{h_1}$ at large $\tan\beta$.

Let us mention that in this scenario with suppressed $h_1f\overline{f}$ and $h_1VV$ couplings, 
$h_1$ can not be produced in the usual channel such as gluon fusions, vector boson fusions or Higgsstrahlung.
According to sum rules Eq.(\ref{sumrule}), if the singlet component of $h_1$ is small and 
$h_1VV$ coupling is suppressed then $h_1ZA$ and $h_1W^\pm H^\mp$ are enhanced, therefore
$h_1$ can be produced in one of the following processes: $pp\to Z^*\to h_1A$ or $pp\to W^*\to h_1H^\pm$
which would lead respectively to the following final states $ZAA$ or $WH^\pm H^\mp$. 

\subsection*{$h_2$ decays}
\begin{figure*}[!h]
\centering
\resizebox{0.32\textwidth}{!}{
\includegraphics{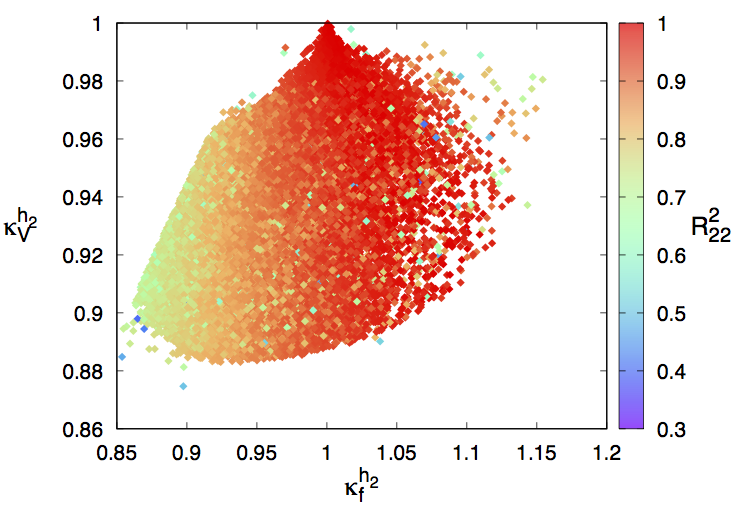}}
\resizebox{0.32\textwidth}{!}{
\includegraphics{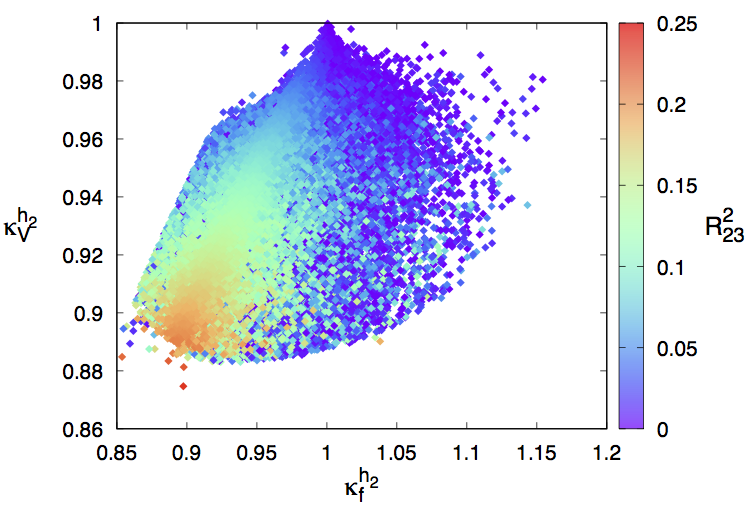}}
\resizebox{0.28\textwidth}{!}{
\includegraphics{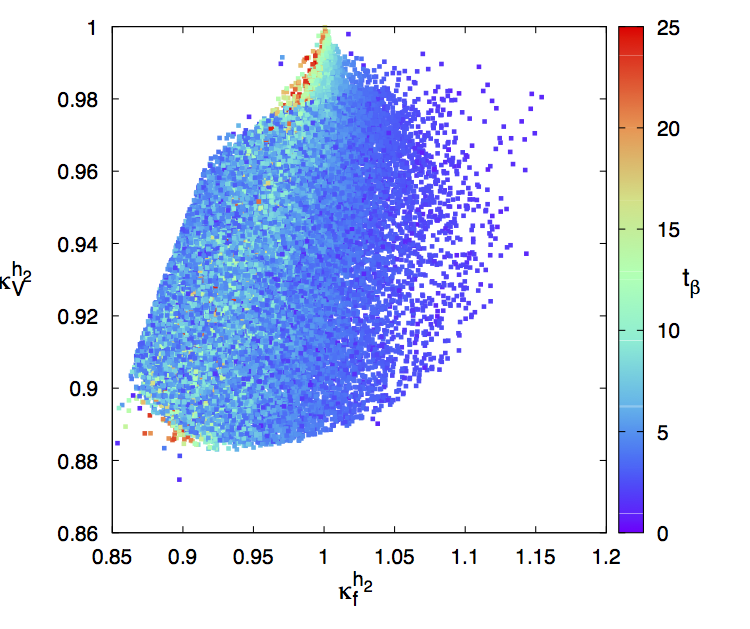}}
\caption{$(\kappa_V^{h_2},\kappa_f^{h_2})$ in 2HDMS type-I as a function of $R_{22,23}$ (left and middle panel), and $\tan\beta$ on the right panel.}
\label{fig:type1-h2-kv-kf}
\end{figure*}

We now discuss the decay of the SM-like $h_2$. We first show the consistency of $h_2\to VV$ and 
$h_2\to f\overline{f}$ with LHC data. For this purpose we illustrate in Figure.(\ref{fig:type1-h2-kv-kf})  
the correlation  between  
$\kappa_V^{h_2}$ and $\kappa_f^{h_2}$ as a function of $R_{2i}^2$.
According to the sum rule Eq.(\ref{sumrulehvv}), $\kappa_V^{h_2}< 1$, and this is clearly 
illustrated in the plot.  One can see from the plot that when $\kappa_V^{h_2}\approx 1$ we have also 
$\kappa_f^{h_2}\approx 1$, this is a consequence of the sum rule Eq.(\ref{sumrulevf}). 
However, the suppression of $\kappa_V^{h_2}$  could be of the order of 12\%
and it could happens both for $\kappa_f^{h_2}<1$ or $\kappa_f^{h_2}>1$. 
Note that the suppression of both $\kappa_f^{h_2}$ and $\kappa_V^{h_2}$ 
takes place when the singlet component of $h_2$ is rather large $R_{23}^2>0.1$.  
One can see that $\kappa_f^{h_2}$ could reach a values less than 0.8 
for $R_{23}^2\approx 0.25$.  It is also clear from the plot that one can have an enhancement of 
$\kappa_f^{h_2}$ in the range of $[1.05-1.15]$ for small singlet component of $h_2$ 
($R_{23}^2\approx 0.1$) and moderate $\tan\beta$. 

\begin{figure*}[!h]
\centering
\resizebox{0.32\textwidth}{!}{
\includegraphics{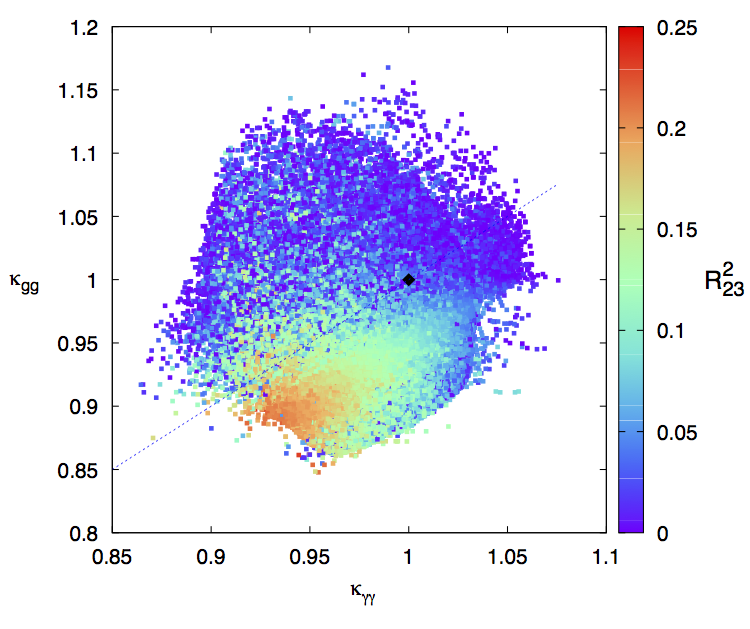}}
\resizebox{0.32\textwidth}{!}{
\includegraphics{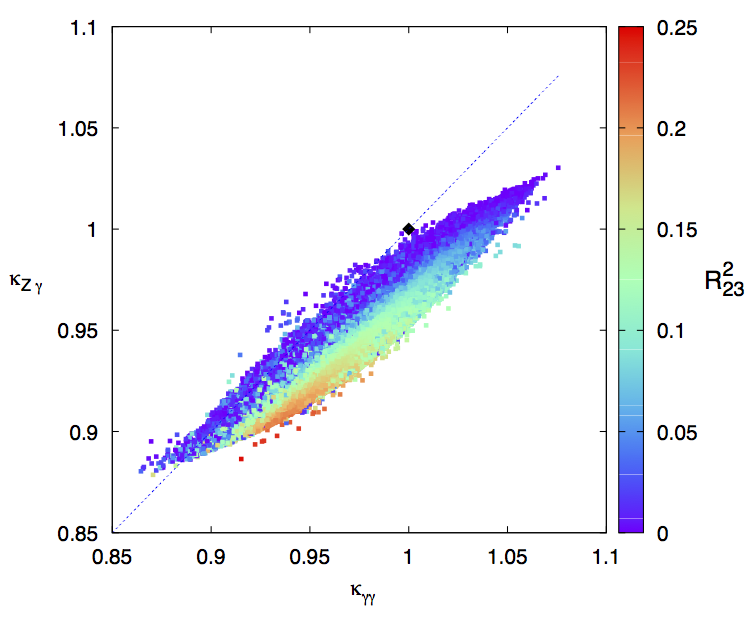}}
\caption{Correlations: between $\kappa_{gg}^{h_2}$ and $\kappa_{\gamma \gamma}^{h_2}$ versus $R_{23}^2$ and 
between $\kappa_{\gamma \gamma}^{h_2}$ and $\kappa_{Z \gamma}^{h_2}$ as a function of the singlet component $R_{23}^2$ in 2HDMS type-I at 95\% $C.L$}
\label{fig:type1-h2-k-g-ga-z}
\end{figure*}

In Figure.(\ref{fig:type1-h2-k-g-ga-z}) we show the correlation between $\kappa_{gg}^{h_2}$ and 
$\kappa_{\gamma\gamma}^{h_2}$ on the left panel and the correlation between 
$\kappa_{\gamma\gamma}^{h_2}$ and $\kappa_{\gamma Z}^{h_2}$ on the right panel. The SM value
is indicated as a black box. One can see that both in   $\kappa_{gg}^{h_2}$,
$\kappa_{\gamma\gamma}^{h_2}$ and $\kappa_{\gamma Z}^{h_2}$ deviation from SM value can reach 15\%.
Note that both in $\kappa_{\gamma\gamma}^{h_2}$ and $\kappa_{\gamma Z}^{h_2}$ one can see that 
in most of the case we have a suppression of the rate compared to SM. The figure shows also that for $h_2$ with 
relatively large singlet component we have suppression of $\kappa_{gg}^{h_2}$,
$\kappa_{\gamma\gamma}^{h_2}$ and $\kappa_{\gamma Z}^{h_2}$ rate. We also stress that most of the cases $\kappa_{\gamma Z}^{h_2}< \kappa_{\gamma \gamma}^{h_2}$ 

\begin{figure*}[!h]
\centering
\resizebox{0.32\textwidth}{!}{
\includegraphics{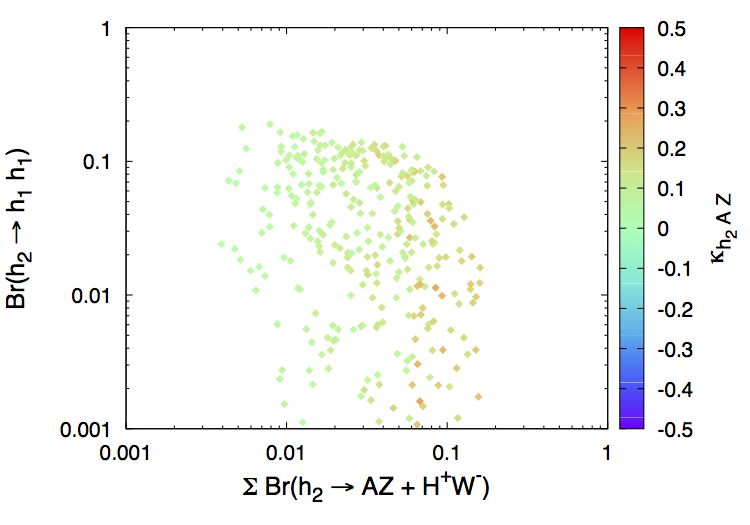}}
\resizebox{0.32\textwidth}{!}{
\includegraphics{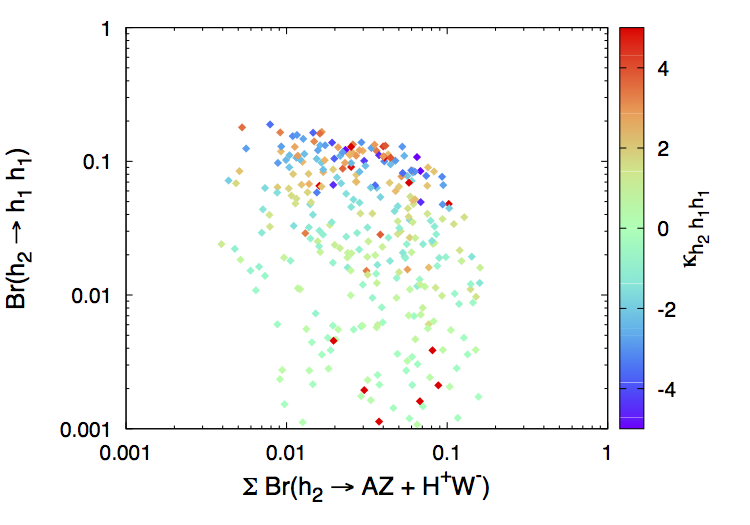}}
\resizebox{0.32\textwidth}{!}{
\includegraphics{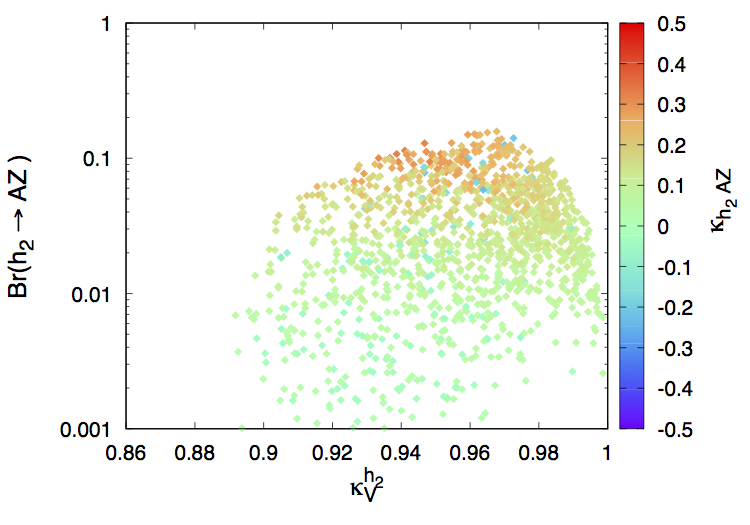}}
\caption{$Br(h_2 \rightarrow h_1 h_1)$ and $Br(h_2 \rightarrow AZ+ H^+W^-)$ versus $\kappa_{h_2 AZ}$ (left) and 
$\kappa_{h_2h_1h_1}$ (middle). On the right panel 
$Br(h_2 \rightarrow AZ)$ versus $\kappa_V^{h_2}$ as a function of $\kappa_{h_2 AZ}$
at 95\% $C.L$ in 2HDMS type-I}
\label{fig:type1-h2-Rij1}
\end{figure*}

As we have seen in Figure.(\ref{fig:type1-h2-kv-kf}), decays of $h_2$  into SM particles such as $WW$, $ZZ$, 
$b\overline{b}$ and $\tau^+\tau^-$ are consistent with LHC measurements with 
deviations from SM predictions  that could goes up to  10-15\%.
However, these deviations are mainly due to experimental uncertainties on all the LHC 
measurements which could be larger than 10\% in some channels. 
Therefore, taking into account these uncertainties,  there is still 
a room for the non-detected SM higgs decays  such as $h_2\to h_1h_1, AA, AA^*, AZ^*, H^\pm W^*$.
In our scan we assume that $m_A>62.5$ GeV, therefore $h_2\to AA$ will not be open and $h_2\to AA^*$ is rather suppressed. We are only left  with 
$h_2\to h_1h_1, AZ^*, H^\pm W^*$ channels. As explained above, all these additional decays of the SM Higgs 
should not exceed 24\% .\\
We show in Figure.(\ref{fig:type1-h2-Rij1})  $Br(h_2\to h_1h_1)$ as a function of 
$Br(h_2\to Z^*A)+Br(h_2\to W^*H^\pm)$ with  $\kappa_{h_2h_1h_1}$ 
on the horizontal axis (left panel). While on the right panel we illustrate the singlet 
component of $h_1$ on the horizontal axis. 
Note that the couplings $h_1AZ$ and  $h_1W^\mp H^\pm$   
are exactly the same (see Eq.(\ref{sumrule})). Therefore, if $m_A\approx m_{H\pm}$
 then $Br(h_2\to Z^*A)$ and $Br(h_2\to W^*H^\pm)$ are of the same order.  
 The total amount for $Br(h_2\to h_1h_1)+Br(h_2\to Z^*A)+Br(h_2\to W^*H^\pm)$ 
 should not exceed 24\% as requested from the non-detected decay of the SM Higgs,
 and this is rather clear from Figure.(\ref{fig:type1-h2-Rij1}). 
 The plots also display the correlation between 
$Br(h_2\to h_1h_1)$  and $Br(h_2\to Z^*A)+Br(h_2\to W^{*\mp}H^\pm)$.
When $Br(h_2\to h_1h_1)$ is maximized, $Br(h_2\to Z^*A)+Br(h_2\to W^{*\mp}H^\pm)$ 
is minimal and vice verse.
One can have also a configuration where both  $Br(h_2\to h_1h_1)$ and 
$Br(h_2\to Z^*A)+Br(h_2\to W^{*\mp}H^\pm)$  are of the same size. 
In the case where both A and $H^\pm$ are heavier than 125 GeV, only $h2\to h_1h_1$ 
would contribute to the non-detected decay of $h_2$.  

 On the middle panel of Figure.(\ref{fig:type1-h2-Rij1})  it is clear that 
 when the reduced coupling of $h_2h_1h_1$ is large, the branching 
 ratio $Br(h_2\to h_1 h_1)$ is substantial which would provide an important 
 production channel for $h_1$ from $h_2$ decay: $gg\to h_2 \to h_1 h_1$
 which could compete with the other production channels such as $pp\to Wh_1$ and/or 
 $pp\to \{h_1A, h_1H^\pm\}$.\\
 On the right panel of Figure.(\ref{fig:type1-h2-Rij1}) we illustrate the correlation between 
 $Br(h_2\to Z^*A)$ and $\kappa_V^{h_2}$ as a function of $\kappa_{h_2AZ}$.
 As one can see from the plot, and according to the sum rule Eq.(\ref{sumrule}), 
 when $h_2VV$ is full strength, then $Br(h_2\to Z^*A)$ is suppressed.

\begin{figure*}[!h]
\centering
\resizebox{0.32\textwidth}{!}{
\includegraphics{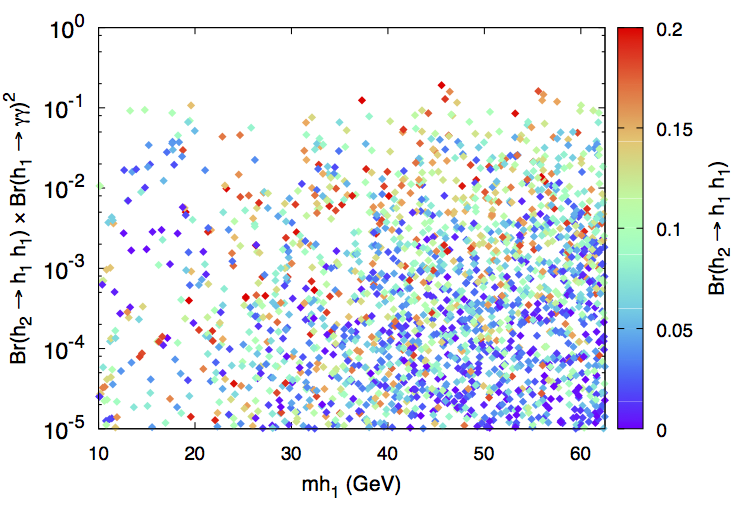}}
\resizebox{0.32\textwidth}{!}{
\includegraphics{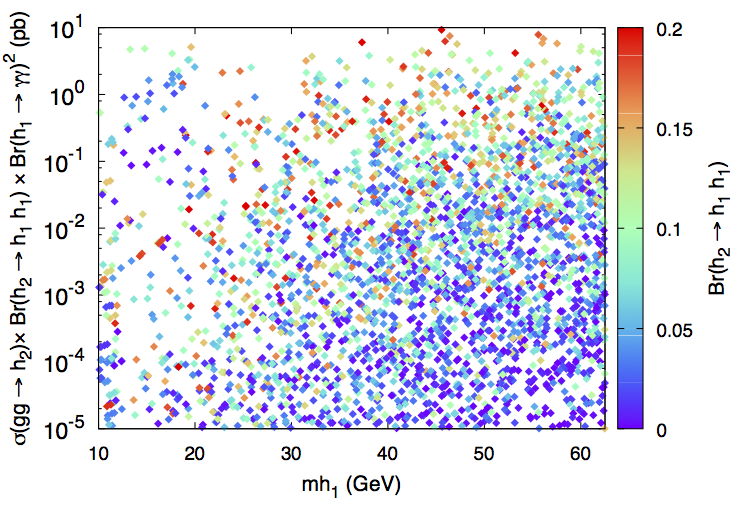}}
\caption{ (Left) $Br(h_2 \rightarrow h_1 h_1)\times Br(h_1 \rightarrow \gamma \gamma)^2$ and  (right) $ \sigma(gg \rightarrow h_2) \times Br(h_2 \rightarrow h_1 h_1)\times Br(h_1 \rightarrow \gamma \gamma)^2$ as a function of $m_{h_1}$ versus $Br(h_2 \rightarrow h_1 h_1)$ at 95\% $C.L$ in 2HDMS type-I}
\label{fig:type1-4gamma}
\end{figure*}

We have seen previously that $Br(h_2\to h_1h_1)$ could be significant and can reach 20\% in some case.
In the case where $h_1$ is dominated by the singlet component, it is well known that 
it is hard to produce it through the conventional channels such as ggF, VBF ect. 
Therefore, the process  $gg\to h_2$ followed by the decay
$h_2\to h_1h_1$ could be an important process for the production of $h_1$. 
In the case where $h_1$ is dominated by the singlet component, its decay to SM 
particle would be suppressed. In such case, it may be possible that $h_1$ would decay 
to a pair of photons which could proceed through charged Higgs loops. 
Therefore, the process $gg\to h_2 \to h_1h_1$ could lead to a spectacular 
4 photons final states.
 In Figure.(\ref{fig:type1-4gamma})(left) we illustrate the branching fraction $Br(h_2\to h_1h_1)\times Br(h_1\to \gamma\gamma)^2$ as a function of $m_{h_1}$. As can be seen, such branching fraction could reach 10\% in some cases.
 On the right panel of Figure.(\ref{fig:type1-4gamma}), we show the production cross section for $\sigma(gg\to h_2)\times Br(h_2\to h_1 h_1) \times Br(h_1\to \gamma \gamma)^2$.
    
We note that for very small singlet component $R_{13}\approx 0$ where $h_1$ is fully  dominated by the doublet components, one could have sizeable $Br(h_2\to h_1h_1)$ as it has been noticed in the usual 2HDM 
\cite{Bernon:2015wef,Arhrib:2017uon}.  

\begin{figure*}[!h]
\centering
\resizebox{0.32\textwidth}{!}{
\includegraphics{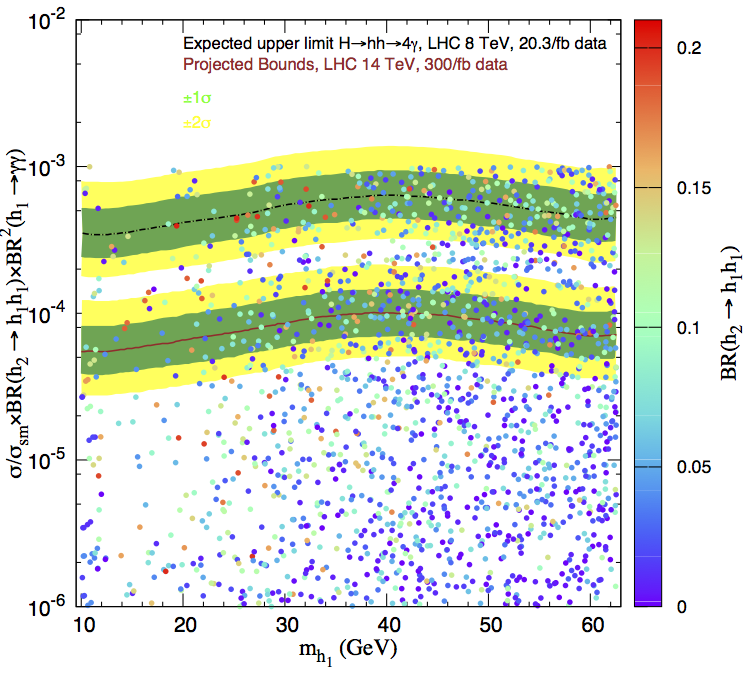}}
\caption{Upper limit at 95\% CL on $\frac{\sigma(h_2)}{\sigma_{sm}}\times Br(h_2 \rightarrow h_1 h_1 \rightarrow 4 \gamma )$ as a function of $mh_1$ vs. $Br(h_2 \rightarrow h_1 h_1)$ from ATLAS searches at 8 TeV (upper band) and the projection for 14 TeV (lower band) taken from \cite{Arhrib:2017uon}. 
The green and yellow color indicate
the allowed regions at 68\% and 95\%, respectively}
\label{atlas4gamma}
\end{figure*}

Recently, ATLAS publish  their results for the  search 
of new phenomena in events with at least three photons
\cite{Aad:2015bua} based on 8 TeV CM energy with 20.3 fb$^{-1}$. 
This search was used to put constraint 
on an N-MSSM scenario  which leads to four photons final states 
$gg \to H \to a_1a_1\to 4 \gamma $ 
where a light pseudo-scalar, 
if dominated by singlet component, can decay fully into two photon with 
100$\%$ branching ratio. Following this work, it has been demonstrated in \cite{Arhrib:2017uon} 
that the kinematic distributions for $qq\to H\to a_1a_1\to 4 \gamma$ 
and $qq\to H\to h_1h_1\to 4 \gamma$ with $h_1$ being CP-even 
are identical. Ref \cite{Arhrib:2017uon}  also provide a projection for 
14 TeV CM energy.
Therefore results from \cite{Aad:2015bua} can be applied 
to our four photons final states.
In Figure.(\ref{atlas4gamma}), we present our predictions for 
$pp\to h_2\to h_1h_1\to 4 \gamma$ both for 8 TeV and 14 TeV
together with the 8 TeV exclusion from ATLAS analysis. 
ATLAS projection for 14 TeV is also shown in the lower band. 
It is clear that some benchmark points are already excluded by the 8TeV data
and the 14 TeV projection. However, several benchmarks are still alive.

\subsection*{$h_3$ decays}
\begin{figure*}[!h]
\centering
\resizebox{0.32\textwidth}{!}{
\includegraphics{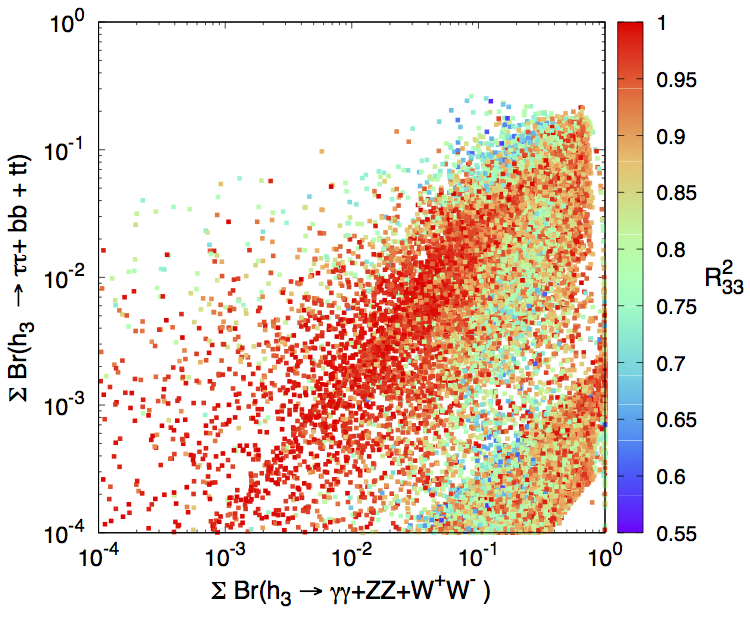}}
\resizebox{0.32\textwidth}{!}{
\includegraphics{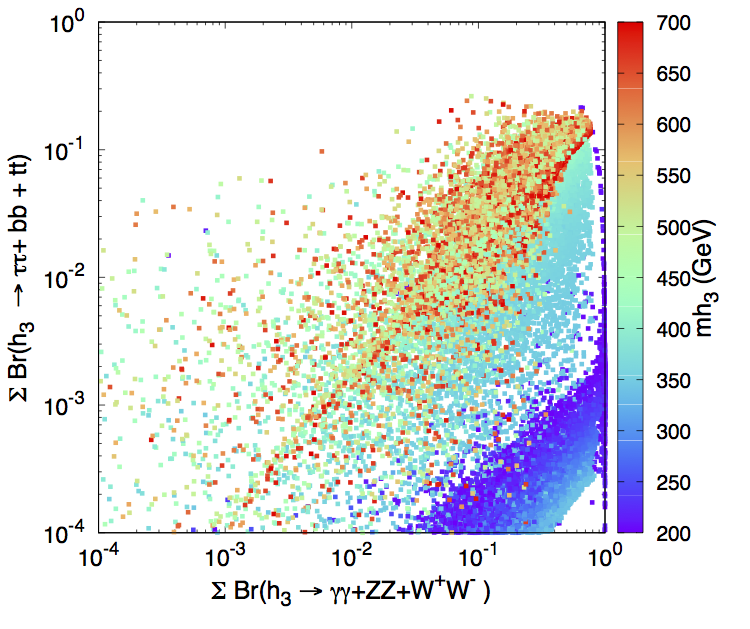}}
\caption{Correlation between $Br(h_3 \rightarrow ff)$ and 
$Br(h_3 \rightarrow VV)$ versus $R_{33}^2$ (left) and 
$m_{h_3}$ (right) (GeV) in 2HDMS type-I}
\label{bF_bV_h3}
\end{figure*}

We now discuss $h_3$ decays. We show in Figure.(\ref{bF_bV_h3}) the branching fractions for $h_3\to ff$, $f=b, \tau, t$ and $h_3\to VV$, $V=\gamma, Z, W$
as a function of singlet component $R_{33}$ and $m_{h_3}$.
It is clear that $h_3$ is dominated by singlet component. One can see that before the opening of the $t\bar{t}$ threshold, $h_3\to WW$ could be the dominant decay mode of $h_3$ with a branching which can reach up to 80\%, while $h_3\to ZZ$ goes up to 20\%  and in such case  $h_3\to h_1 h_1$ is suppressed.
After opening of $t\bar{t}$ threshold, $h_3 \to t\bar{t}$ can be slightly larger than  10\% for large $m_{h_3}$ mass.

\begin{figure*}[!h]
\centering
\resizebox{0.32\textwidth}{!}{
\includegraphics{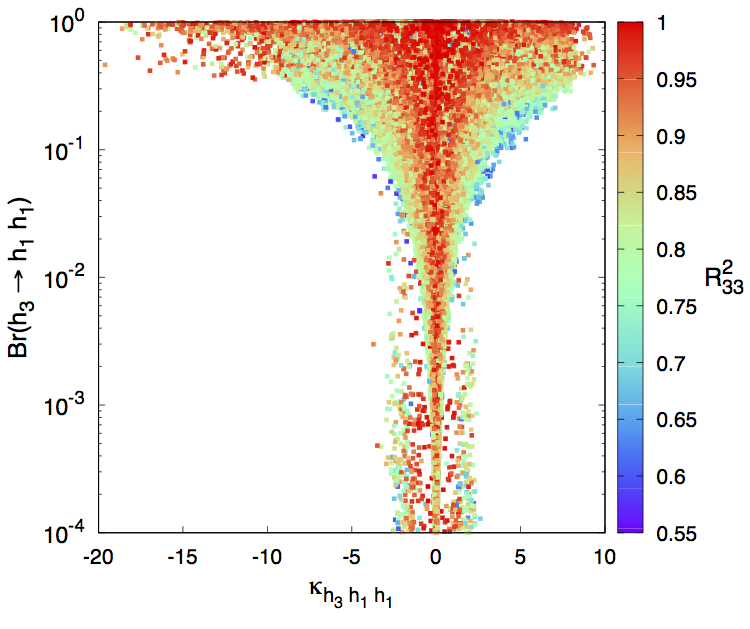}}
\resizebox{0.32\textwidth}{!}{
\includegraphics{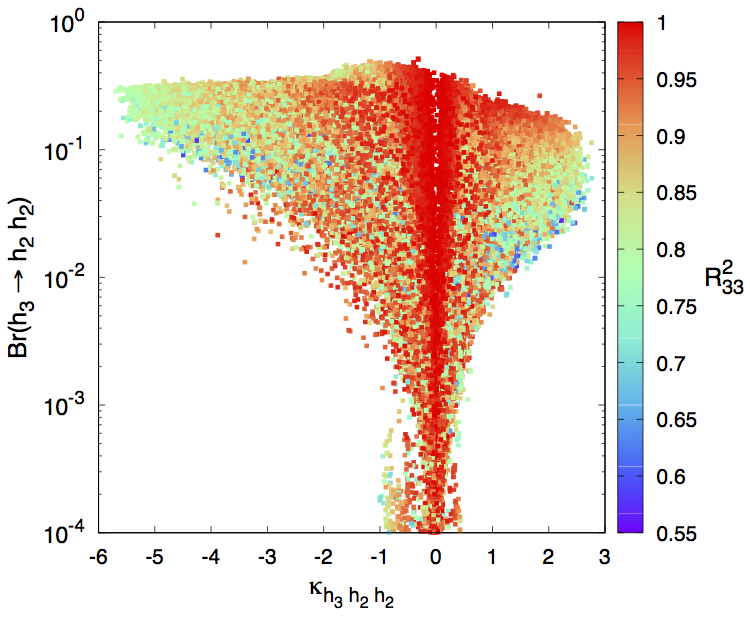}}
\resizebox{0.32\textwidth}{!}{
\includegraphics{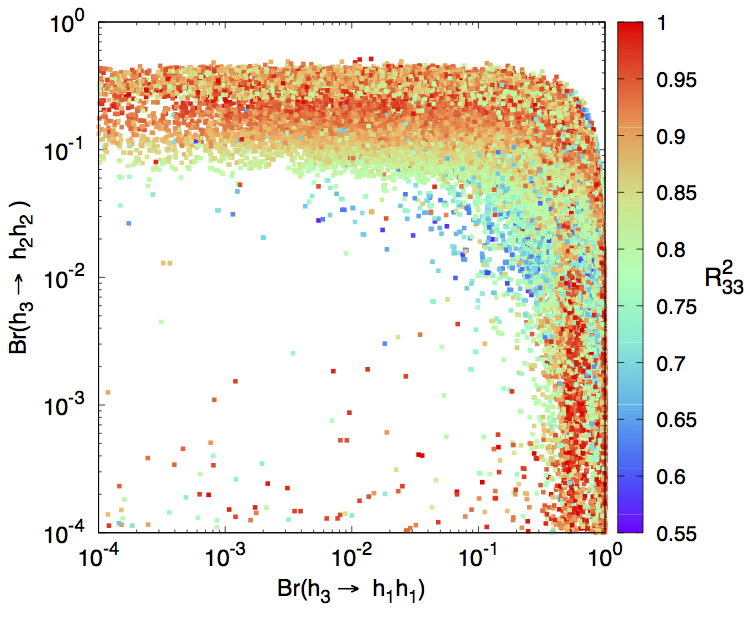}}\\
\resizebox{0.32\textwidth}{!}{
\includegraphics{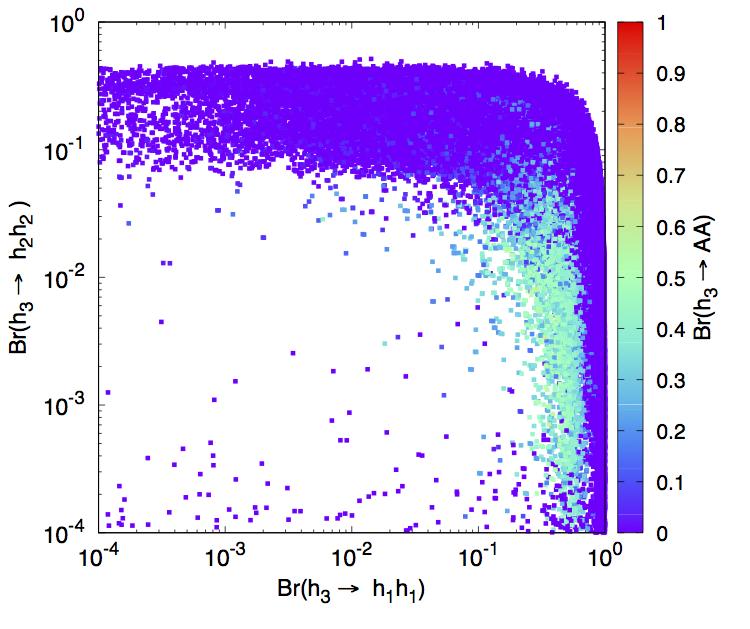}}
\resizebox{0.32\textwidth}{!}{
\includegraphics{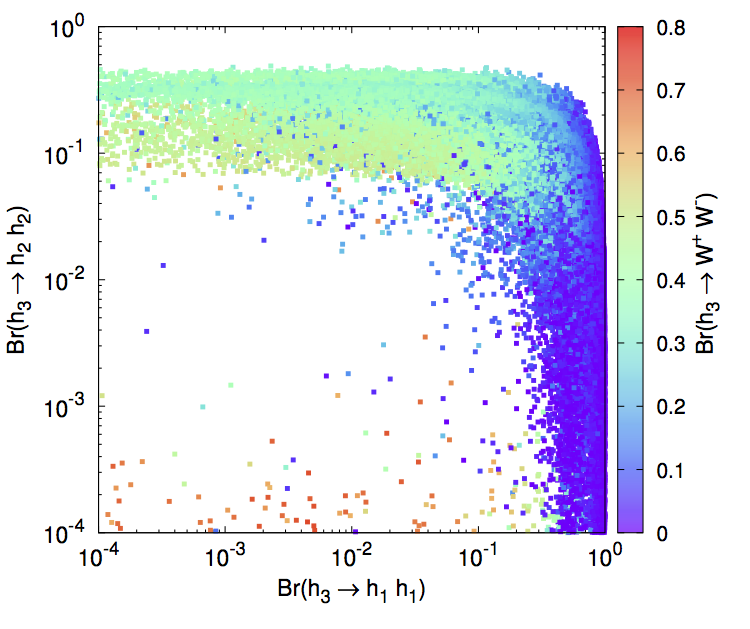}}
\resizebox{0.32\textwidth}{!}{
\includegraphics{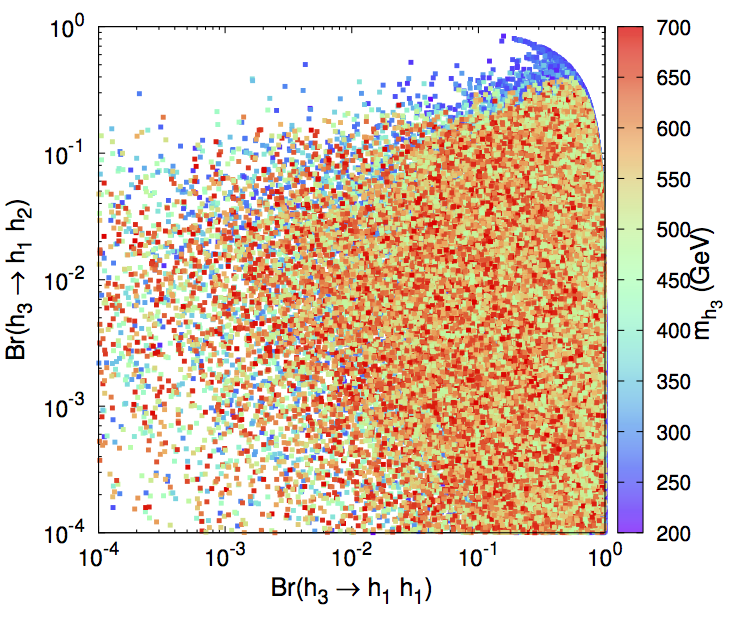}}
\caption{Upper panels: (left) $Br(h_3\to h_1 h_1)$ as a function of  
$\kappa_{h_3 h_1 h_1}$ and (middle) $Br(h_3\to h_2 h_2)$ as a function of 
 $\kappa_{h_3 h_2 h_2}$ with  $R_{33}^2$ displayed on the vertical and (right) the correlation between $Br(h_3\to h_1 h_1)$ and $Br(h_3\to h_2 h_2)$ .
 Lower panels: correlation between $Br(h_3\to h_1 h_1)$ and $Br(h_3\to h_2 h_2)$ 
 as a function of $Br(h_3\to AA)$ (lower-left),  as a function of $Br(h_3\to WW)$  (lower-middle) 
  and correlation between $Br(h_3\to h_1 h_1)$ and $Br(h_3\to h_1 h_2)$ as a function of $m_{h_3}$.}
\label{h3_tril}
\end{figure*}

We now discuss higgs to higgs decays, such as $h_3\to h_1h_1,h_1h_2, h_2h_2$ and 
$h_3\to ZA, W^\pm H^\mp$. In Figure.(\ref{h3_tril}) (upper plot) we illustrate 
the branching ratio of $h_3\to h_1h_1$ (left), $h_3\to h_2h_2$ (middle) 
and their correlation (right).
From the left panel, one can see that Br$(h_3\to h_1h_1)$ can be substantial and 
becomes the dominant decay mode, while from the middle panel it is clear that  
 Br$(h_3\to h_2h_2)$ can reach 30\% as a maximal value.
 In the case where $Br(h_3\to h_1h_1)$ is the dominant decay, 
 then one can have a new production mechanism for $h_1$, 
 namely: $pp\to h_3\to h_1 h_1$. This production channel might be useful for the case where $h_1$ 
 has large singlet component in which case it will be challenging to produce it in the conventional channels.\\ 
In the lower panels of  Figure.(\ref{h3_tril}) we display the correlation between 
$Br(h_3\to h_1 h_1)$, $Br(h_3\to h_2 h_2)$ , $Br(h_3\to AA)$ , $Br(h_3\to h_1h_2)$ as well as with $Br(h_3\to WW)$. \\
It is clear that one can  have a scenario where both $Br(h_3\to h_2 h_2)$
and $Br(h_3\to AA)$ are rather large with branching fractions of the order 40\%. 
It is also clear that when $Br(h_3\to h_1 h_1)$ and $Br(h_3\to h_2 h_2)$ are suppressed then  
$Br(h_3\to WW, ZZ)$ would become slightly large.

\begin{figure*}[!h]
\centering
\resizebox{0.32\textwidth}{!}{
\includegraphics{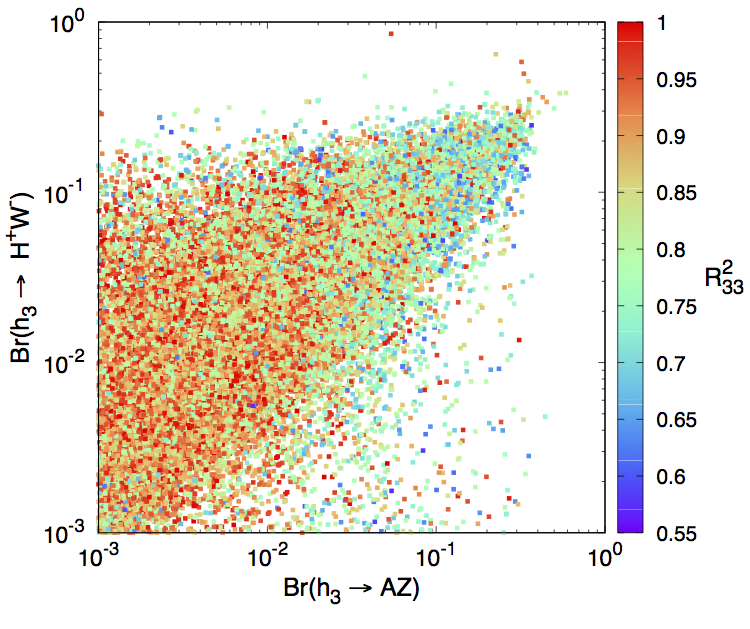}}
\resizebox{0.32\textwidth}{!}{
\includegraphics{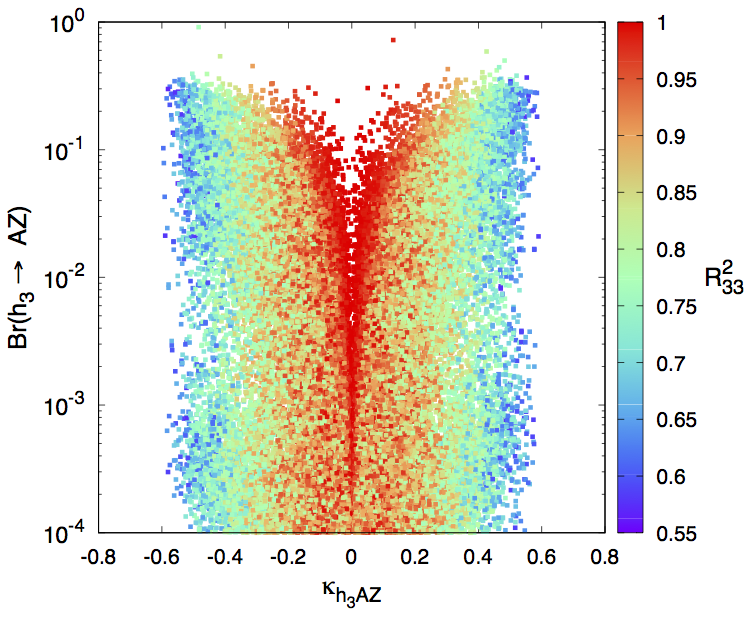}}
\caption{Correlation between $Br(h_3 \rightarrow AZ)$ and $Br(h_3 \rightarrow H^\pm W^\mp)$ versus $R_{33}^2$ (left panel) and $Br(h_3 \rightarrow AZ)$ as a function of $\kappa_{h_3 AZ}$ and $R_{33}^2$(right panel) }
\label{h3SV_decay}
\end{figure*}

In Figure.(\ref{h3SV_decay}) we show the branching fractions of 
$h_3 \rightarrow AZ$ and $h_3 \rightarrow H^\pm W^\mp$ versus 
$R_{33}^2$. As one can see from the plots and according to the sum-rule
Eq.(\ref{sumrule}), when $h_3$ is dominated by the singlet component 
$R_{33}\approx 1$, then both $h_3ZA$ and $h_3W^\mp H^\pm$ couplings  
are suppressed  resulting in a small branching ratio for both channels. 
For $R_{33}$ away from 1, the branching fraction $h_3 \rightarrow AZ$ and 
$h_3 \rightarrow H^\pm W^\mp$ can be in the range of 10-40\% 
 in some cases.

\begin{figure*}[!h]
\centering
\resizebox{0.32\textwidth}{!}{
\includegraphics{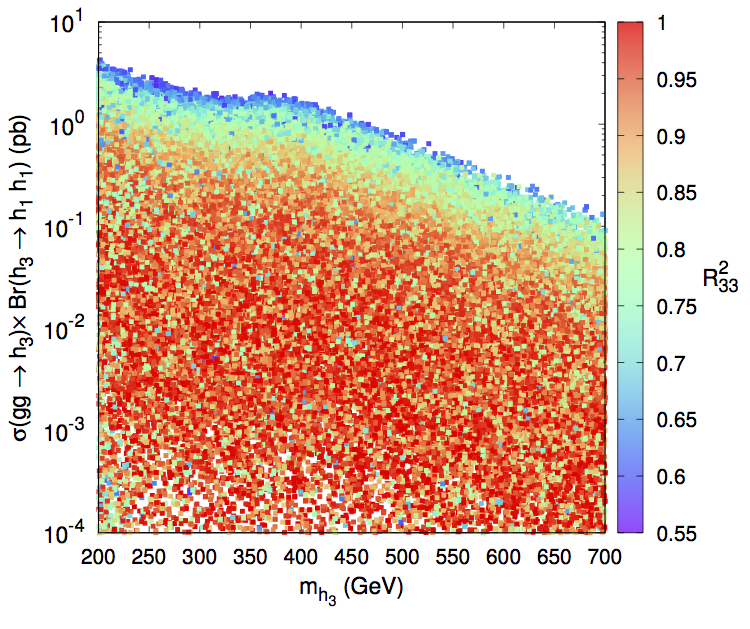}}
\resizebox{0.32\textwidth}{!}{
\includegraphics{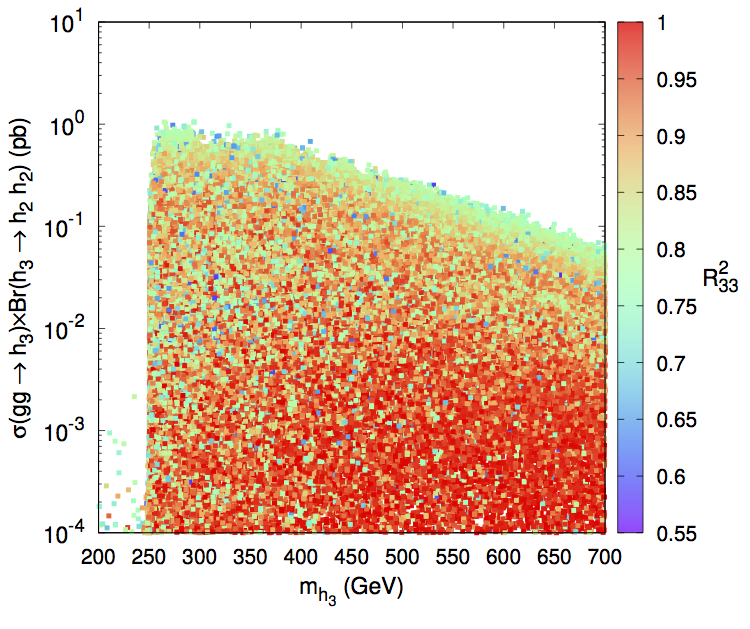}}
\resizebox{0.32\textwidth}{!}{
\includegraphics{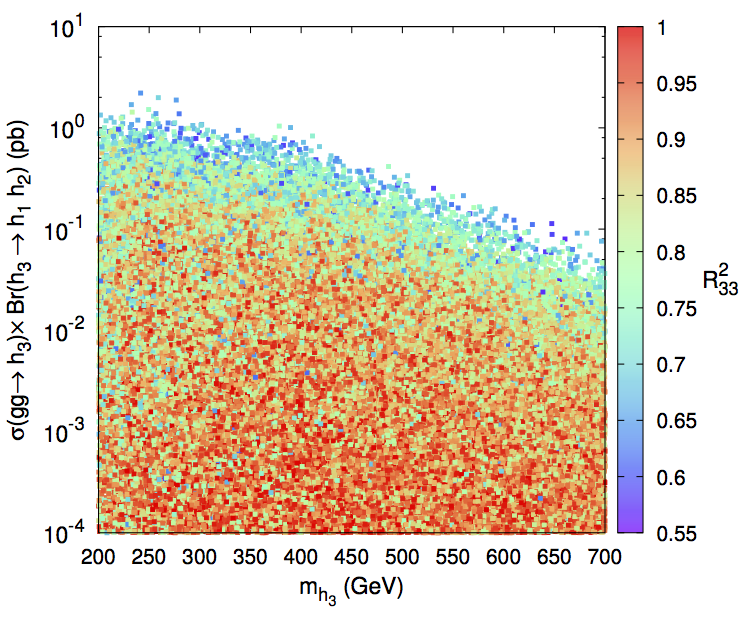}}
\caption{$\sigma(pp \rightarrow h_3)\times Br(h_3 \rightarrow h_2h_2)$(left panel), $\sigma(pp \rightarrow h_3)\times Br(h_3 \rightarrow h_1h_1)$ (middle panel) and $\sigma(pp \rightarrow h_3)\times Br(h_3 \rightarrow h_1h_2)$ (right panel) as a function of $m_{h_3} (GeV)$ and $R_{33}^2$ in 2HDMS type-I}
\label{sec_h3_SS}
\end{figure*}

Given that $Br(h_3\to h_1h_1)$ can be sizeable, one can look to the amount of production cross section 
that one can get from $h_3$ production followed by 
$h_3$ decay into a pair of $h_1$. We illustrate in 
Figure.(\ref{sec_h3_SS}) both $pp\to h3\to h_2h_2$ (left), $pp\to h_3\to h_1h_1$ (middle) and 
 $pp\to h_3\to h_1h_2$ (right).
One can see  that the production rate of $h_1$ is large especially in the mass range $200\ \text{GeV}\lesssim m_{h_3}\lesssim 250\ \text{GeV}$  when the decay channel $h_3 \rightarrow h_1 h_1$ is kinematically allowed 
 and both $h_3 \rightarrow h_2 h_2$ and $h_3 \rightarrow t \overline{t}$ are closed. 
 The same behavior is observed in the middle panel, where the magnitude in the cross section is larger in the mass range $250\ \text{GeV}\lesssim m_{h_3}\lesssim 350\ \text{GeV}$ when the decay channel $h_3 \rightarrow h_2 h_2$ is opened and $h_3 \rightarrow t \overline{t}$ mode is closed.

\section{Conclusions}
We have studied the two Higgs doublet model extended with a real singlet scalar.
The spectrum contains 3 CP-even $h_{1,2,3}$, one CP-odd and a pair of charged Higgs. We derive full set of perturbative unitarity constraints, boundedness from below constraints as well as the oblique parameters S, T and U.\\
In our analysis, we concentrate on the case where
$h_2$ is the SM Higgs like particle observed at the LHC and assume that 
$h_1$ is lighter than 125 GeV. We study the consistency of our scenario with both LHC data taken at 
8 TeV and 13 TeV as well as with all the available LEP-II and Tevatron data. We have shown in the framework of 2HDMS that:
\begin{itemize}
    \item $h_1$ can be quasi-fermiophobic and would decay dominantly into two photons.
	\item LHC data still allow a room for the non-detected decays of the SM-Higgs
	$h_2\to h_1 h_1$ and others with a branching ratio of the order which can reach 24\%. Such decay followed by two photons decay of $h_1$ could lead to four photons signature, namely $pp\to h_2\to h_1 h_1\to 4 \gamma$.
	\item Comparison of ATLAS data with our four photons signal show that 
	there is a large area of parameter space that still escapes ATLAS data
\end{itemize}
We have also shown that in 2HDMS type I, $h_2$ and $h_3$ can decay to some exotic modes such as 
$h_{2,3}\to h_1h_1$, $h_{2,3}\to Z A$ and $h_{2,3}\to W \pm H^\mp$ with substantial branching ratio.
The production process $gg\to h_{2,3}$ followed by the decays $h_{2,3}\to h_1h_1, h_1h_2$ 
could be sizeable and could be an important source of production of  $h_1$ in the case where $h_1$ have a large singlet component
where it is rather difficult to produce it using the conventional channel.

\section*{Acknowledgments}
This work is supported by the Moroccan Ministry of Higher Education and Scientific Research MESRSFC and  CNRST: Projet PPR/2015/6. 

\section*{Appendix}
\appendix
\setcounter{equation}{0}
\renewcommand{\theequation}{A\arabic{equation}}
\section{BFB constraints}
\label{appendix-bfb}

Consider for instance the following case in  which there is no coupling
between doublets $H_{i}$ and singlet $S$ Higgs bosons, i.e. 
$\lambda_3 = \lambda_4 = \lambda_5 = \lambda_7= \lambda_8=0$,  it is obvious that 
\begin{equation}
\lambda_1 >0 \;\;{\rm \&}\;\; \lambda_2 >0 \;\;{\rm \&}\;\; \lambda_6 >0 
\end{equation}
\noindent
To proof the necessary and sufficient BFB conditions, we adopt a different parameterization 
of the fields that will turn out to be particularly convenient to entirely solve the problem. 
For that we define :
\begin{eqnarray}
r &\equiv& \sqrt{H_1^\dagger{H_1} + H_2^\dagger{H_2} + S^\dagger{S}} \\
H_1^\dagger{H_1} &\equiv& r^2 \cos^2 \theta \sin^2 \phi \\
H_2^\dagger{H_2} &\equiv& r^2 \sin^2 \theta \sin^2 \phi  \\
S^\dagger{S} &\equiv& r^2 \cos^2 \phi  \\
\frac{H_1^\dagger\,H_2}{|H_1||H_2|} &\equiv&  \xi e^{i\psi} 
\end{eqnarray}.
\noindent
Obviously, when $H_1$, $H_2$ and $S$ scan all the field space, the radius $r$ 
scans the domain $[0, \infty[$, the angle $\theta \in [0, 2 \pi]$ and the angle 
$\phi \in [0, \frac{\pi}{2}]$. Moreover, one can show that $\frac{H_1^\dagger}{|H_1|}\cdot\frac{H_2}{|H_2|}$ 
is a product of unit spinor, so that $\xi \in [0,1]$.\\

\noindent
With this parameterization, one can cast $V^{(4)}(H_1,H_2,S)$ in the following simple form,
\begin{widetext}
\begin{eqnarray}
V^{(4)}(r,c^2_\theta,s^2_\phi,c_{2\psi},\xi) &=& r^4\Big\{ \frac{\lambda_1}{2}c^4_\theta s^4_\phi + \frac{\lambda_2}{2}s^4_\theta s^4_\phi + \lambda_3 c^2_\theta s^2_\theta s^4_\phi + \lambda_4 c^2_\theta s^2_\theta s^4_\phi \xi^2 \nonumber\\
&& + \frac{\lambda_5}{2}c^2_\theta s^2_\theta s^4_\phi \xi^2 (e^{2i\psi}+e^{-2i\psi})  + \frac{\lambda_6}{8}c^4_\phi + \frac{1}{2} (\lambda_7 c^2_\theta s^2_\phi c^2_\phi + \lambda_8 s^2_\theta s^2_\phi c^2_\phi) \Big\}\nonumber\\
\label{eq:V4general2}
\end{eqnarray}
\end{widetext}
\noindent
Let define again :
\begin{eqnarray}
 x &\equiv& \cos^2\theta \quad , \quad y \equiv \sin^2\phi  \quad , \quad
 z \equiv  \cos2\psi
\end{eqnarray}
which allows to write 
\begin{widetext}
\begin{eqnarray}
V^{(4)}(x,y,z,\xi) &=& \Big\{\frac{\lambda_1}{2}\,x^2 + \frac{\lambda_2}{2}\,(1-x)^2 + \lambda_3\,x(1-x) + \lambda_4\,x(1-x)\xi^2 + \lambda_5\,x(1-x)\xi^2\,z \Big\}\,y^2\nonumber\\
&& + \frac{\lambda_6}{8}\,(1-y)^2\nonumber\\
&& + \Big\{ \frac{\lambda_7}{2}\,x + \frac{\lambda_8}{2}\,(1-x)\Big\}\,y(1-y)
\label{eq:V4general3}
\end{eqnarray}
\end{widetext}
it is easy to find the constraint conditions by studying $V^{(4)}(x,y,z,\xi) $ as a quadratic function using the fact that :
\begin{widetext}
\begin{eqnarray}
f(\zeta) =  a\,\zeta^2 + b\,(1-\zeta)^2 + c\,\zeta\,(1 - \zeta),\quad \zeta \in (0,1)\quad \Leftrightarrow \quad a > 0,\, b > 0\,\,{\rm and}\,\,c +2\sqrt{ab} > 0\label{fzeta}\nonumber\\
\end{eqnarray}
\end{widetext}
we can deduce the set of constraints as :
\begin{eqnarray}
 A &\equiv& \frac{\lambda_1}{2}\,x^2 + \frac{\lambda_2}{2}\,(1-x)^2 + \lambda_3\,x(1-x) +\nonumber\\
 && \lambda_4\,x(1-x)\xi^2 + \lambda_5\,x(1-x)\xi^2\,z > 0 \label{condA}\\
 B &\equiv& \frac{\lambda_6}{8} > 0 \label{condB}  \\
 C &\equiv& \frac{\lambda_7}{2}\,x + \frac{\lambda_8}{2}\,(1-x) > -2\sqrt{A\,B} \label{condC}
\end{eqnarray}
the simple condition can be extracted from Eq.(\ref{condB}), which imply $\lambda_6 > 0$.
For $A > 0$ one can use Eq.(\ref{fzeta}) again to get the ordinary 2${\rm HDM}$ BFB constraints taking into account if $\xi = {0;1}$ and $z={-1;1}$ :\\
\begin{eqnarray}
 \lambda_1\,,\,\lambda_2 &>& 0 \label{2hdm_1}\\
 \lambda_3 + \sqrt{\lambda_1\lambda_2} &>& 0 \label{2hdm_2}  \\
 \lambda_3 + \lambda_4 + |\lambda_5| + \sqrt{\lambda_1\lambda_2} &>& 0 \label{2hdm_3} 
\end{eqnarray}

For the Eq.(\ref{condC}), we can consider two scenarios :\\ 
\begin{itemize}
\item scenario (1): $\lambda_7$ and $\lambda_8$ $> 0$
\item[] starting with the fact that $x = \cos^2\theta > 0$ and $1-x =\sin^2\theta > 0$, thus $C > 0 \rightarrow \lambda_7,\lambda_8 > 0$ imply that $AB > 0$ which already done in Eqs. (\ref{condA}-\ref{condB}).
\item scenario (2): $\lambda_7$ or $\lambda_8$ $< 0$ 
\item[] this scenario implies $\lambda_7\,{\rm or\backslash and}\,\lambda_8 \le 0$ and it leads to two cases :
\end{itemize}
\begin{eqnarray}
C > -2\sqrt{AB} \quad \Leftrightarrow \quad   \left\{
\begin{aligned}
-2&\sqrt{AB} < C < 2 \sqrt{AB}  \ \ &\text(i)\\   
&&\text{or}\\
C &> 2\sqrt{AB}  \ \ &\text(ii)\\
\end{aligned}
\right.\nonumber\\
\end{eqnarray}

For scenario(i), we can rewrite it like this:
\begin{widetext}
\begin{eqnarray}
(\lambda_1\lambda_6 - \lambda_7^2)\,x^2 + (\lambda_2\lambda_6 - \lambda_8^2)\,(1-x)^2 + 2 \big\{ \lambda_6\,(\lambda_3+\lambda_4\xi^2+\lambda_5\xi^2\,z) - \lambda_7\lambda_8 \big\}x(1-x)&>& 0\nonumber\\
\end{eqnarray}
\end{widetext}
by applying the {\it lemme} given by Eq.(\ref{fzeta}), we get the following new constraints :
\begin{widetext}
\begin{eqnarray}
&& \lambda_7^2 < \lambda_1\lambda_6 \label{2hdm1S_1}  \\ 
&& \lambda_8^2 < \lambda_2\lambda_6 \label{2hdm1S_2}\\
&&\lambda_6\,(\lambda_3+\lambda_4\xi^2+\lambda_5\xi^2\,z) - \lambda_7\lambda_8 + \sqrt{(\lambda_1\lambda_6 - \lambda_7^2)(\lambda_2\lambda_6 - \lambda_8^2)} > 0 \label{2hdm1S_3}  
\end{eqnarray}
\end{widetext}
 
Both Eq.(\ref{2hdm1S_1}) and Eq.(\ref{2hdm1S_2}) gives constraints on $\lambda_7$ and $\lambda_8$: 
\begin{eqnarray}
\hspace{-0.5cm}
-\sqrt{\lambda_1\lambda_6}\,(-\sqrt{\lambda_2\lambda_6})< \lambda_7\,(\lambda_8) < \sqrt{\lambda_1\lambda_6}\,(\sqrt{\lambda_2\lambda_6} )
\label{l7_8} 
\end{eqnarray}
since $\xi = {0;1}$ and $z={-1;1}$, the latest equations leads to
\begin{eqnarray}
&&\hspace{-0.2cm} \lambda_3\,\lambda_6 - \lambda_7\lambda_8 + \sqrt{(\lambda_1\lambda_6 - \lambda_7^2)(\lambda_2\lambda_6 - \lambda_8^2)} > 0 \label{2hdm1S_4} \\
&&\hspace{-0.2cm} \lambda_6\,(\lambda_3+\lambda_4 +|\lambda_5|) - \lambda_7\lambda_8 + \sqrt{(\lambda_1\lambda_6 - \lambda_7^2)(\lambda_2\lambda_6 - \lambda_8^2)} > 0 \nonumber\\\label{2hdm1S_5} 
\end{eqnarray}  
In the other hand, scenario(ii) leads to a contradiction because it implies that $\lambda_7>0$ and $\lambda_8 > 0$ , which is not our case.
%

\setcounter{equation}{0}
\renewcommand{\theequation}{B\arabic{equation}}
\section{Unitarity constraints}
\label{appendix-unitarity}

The first submatrix ${\cal M}_1$  corresponds to scattering whose
initial and final states are one of the following :
$(\phi_1^+\phi_2^-$,$\phi_2^+\phi_1^-$, $\phi_1\chi_2$,$\phi_2\chi_1$,$\phi_s\chi_1$,$\phi_s\chi_2$,$\phi_1\phi_2$,$\phi_1\phi_s$,$\phi_2\phi_s$,$\chi_1\chi_2)$. We have to write out the full matrix, one finds,
%
\begin{widetext}
\begin{eqnarray}
{\cal M}_1 = \left(
 \begin{array}{cccccccccccc}
\displaystyle \lambda_{34}^+ \, & \, \displaystyle 2\lambda_5 \, & \,  \displaystyle -i\frac{\lambda_{45}^-}{2} \, & \, \displaystyle +i\frac{\lambda_{45}^-}{2} \, & \, \displaystyle 0 \, & \, \displaystyle 0 \, & \, \displaystyle \frac{\lambda_{45}^+}{2} \, & \, \displaystyle 0 \, & \, \displaystyle 0 \, & \, \displaystyle \frac{\lambda_{45}^+}{2}  \\
\displaystyle 2\lambda_5 \,&\, \displaystyle \lambda_{34}^+ \,&\,  \displaystyle +i\frac{\lambda_{45}^-}{2} \,&\, \displaystyle -i\frac{\lambda_{45}^-}{2} \,&\, \displaystyle 0 \,&\, \displaystyle 0 \,&\, \displaystyle \frac{\lambda_{45}^+}{2} \,&\, \displaystyle 0 \,&\, \displaystyle 0 \,&\, \displaystyle \frac{\lambda_{45}^+}{2}  \\
\displaystyle +i\frac{\lambda_{45}^-}{2} \,&\, \displaystyle -i\frac{\lambda_{45}^-}{2} \,&\,  \displaystyle \lambda_L \,&\, \displaystyle \lambda_5 \,&\, \displaystyle 0 \,&\, \displaystyle 0 \,&\, \displaystyle 0 \,&\, \displaystyle 0 \,&\, \displaystyle 0 \,&\, \displaystyle 0  \\
\displaystyle -i\frac{\lambda_{45}^-}{2} \,&\, \displaystyle +i\frac{\lambda_{45}^-}{2} \,&\,  \displaystyle \lambda_5 \,&\, \displaystyle \lambda_L \,&\, \displaystyle 0 \,&\, \displaystyle 0 \,&\, \displaystyle 0 \,&\, \displaystyle 0 \,&\, \displaystyle 0 \,&\, \displaystyle 0  \\
\displaystyle 0 \,&\, \displaystyle 0 \,&\,  \displaystyle 0 \,&\, \displaystyle 0 \,&\, \displaystyle \frac{\lambda_7}{2} \,&\, \displaystyle 0 \,&\, \displaystyle 0 \,&\, \displaystyle 0 \,&\, \displaystyle 0 \,&\, \displaystyle 0  \\
\displaystyle 0 \,&\, \displaystyle 0 \,&\,  \displaystyle 0 \,&\, \displaystyle 0 \,&\, \displaystyle 0 \,&\, \displaystyle \frac{\lambda_8}{2} \,&\, \displaystyle 0 \,&\, \displaystyle 0 \,&\, \displaystyle 0 \,&\, \displaystyle 0  \\
\displaystyle \frac{\lambda_{45}^+}{2} \,&\, \displaystyle \frac{\lambda_{45}^+}{2} \,&\,  \displaystyle 0 \,&\, \displaystyle 0 \,&\, \displaystyle 0 \,&\, \displaystyle 0 \,&\, \displaystyle \lambda \,&\, \displaystyle 0 \,&\, \displaystyle 0 \,&\, \displaystyle \lambda_5 \\
\displaystyle 0 \,&\, \displaystyle 0 \,&\,  \displaystyle 0 \,&\, \displaystyle 0 \,&\, \displaystyle 0 \,&\, \displaystyle 0 \,&\, \displaystyle 0 \,&\, \displaystyle \frac{\lambda_7}{2} \,&\, \displaystyle 0 \,&\, \displaystyle 0  \\
\displaystyle 0 \,&\, \displaystyle 0 \,&\,  \displaystyle 0 \,&\, \displaystyle 0 \,&\, \displaystyle 0 \,&\, \displaystyle 0 \,&\, \displaystyle 0 \,&\, \displaystyle 0 \,&\, \displaystyle \frac{\lambda_8}{2} \,&\, \displaystyle 0  \\
\displaystyle \frac{\lambda_{45}^+}{2} \,&\, \displaystyle \frac{\lambda_{45}^+}{2} \,&\,  \displaystyle 0 \,&\, \displaystyle 0 \,&\, \displaystyle 0 \,&\, \displaystyle 0 \,&\, \displaystyle \lambda_5 \,&\, \displaystyle 0 \,&\, \displaystyle 0 \,&\, \displaystyle \lambda 
\end{array}\right) 
\end{eqnarray}
\end{widetext}
where $\lambda=\lambda_3+\lambda_4+\lambda_5$ and $\lambda_L=\lambda_3+\lambda_4-\lambda_5$. We find that ${\cal M}_1$ has the following eigenvalues given by :
\begin{eqnarray}
& & a_{1} =  \lambda_3+\lambda_4 \\
& & a_{2} = \lambda_3-\lambda_5 \\
& & a_{3} = \lambda_3+\lambda_5 \\
& & a_{4} = \frac{\lambda_7}{2}  \\
& & a_{5} = \frac{\lambda_8}{2} \\
& & a_{\pm} =  \lambda_3+2\lambda_4 \pm 3\lambda_5
\end{eqnarray}
The second submatrix ${\cal M}_2$ corresponds to scattering with one of the following
initial and final states:
$(\phi_1^+\phi_1^-$, $\phi_2^+\phi_2^-$, $\frac{\phi_1\phi_1}{\sqrt{2}}$,$\frac{\phi_2\phi_2}{\sqrt{2}}$,$\frac{\phi_s\phi_s}{\sqrt{2}}$,$\frac{\chi_1\chi_1}{\sqrt{2}}$,$\frac{\chi_2\chi_2}{\sqrt{2}})$, where the $\sqrt{2}$ accounts for
identical particle statistics. One finds that ${\cal M}_2$ is given by:

\begin{eqnarray}
{\cal M}_2 = \left(
\begin{array}{ccccccc}
\displaystyle\, 2\lambda_1 \,&\, \displaystyle\, \lambda_{34}^+ \,&\, \displaystyle\, \frac{\lambda_1}{\sqrt{2}} \,&\, \displaystyle\, \frac{\lambda_3}{\sqrt{2}} \,&\, \displaystyle\, \frac{\lambda_7}{2\sqrt{2}} \,&\, \displaystyle\, \frac{\lambda_1}{\sqrt{2}} \,&\, \displaystyle\, \frac{\lambda_3}{\sqrt{2}} \\
\displaystyle\, \lambda_{34}^+ \,&\, \displaystyle\, 2\lambda_2 \,&\, \displaystyle\, \frac{\lambda_3}{\sqrt{2}} \,&\, \displaystyle\, \frac{\lambda_2}{\sqrt{2}} \,&\, \displaystyle\, \frac{\lambda_8}{2\sqrt{2}} \,&\, \displaystyle\, \frac{\lambda_3}{\sqrt{2}} \,&\, \displaystyle\, \frac{\lambda_2}{\sqrt{2}} \\
\displaystyle\, \frac{\lambda_1}{\sqrt{2}} \,&\, \displaystyle\, \frac{\lambda_3}{\sqrt{2}}  \,&\, \displaystyle\, \frac{3\lambda_1}{2} \,&\, \displaystyle\, \frac{\lambda}{2} \,&\, \displaystyle\, \frac{\lambda_7}{4} \,&\, \displaystyle\, \frac{\lambda_1}{2} \,&\, \displaystyle\, \frac{\lambda_L}{2} \\
\displaystyle\, \frac{\lambda_3}{\sqrt{2}} \,&\, \displaystyle\, \frac{\lambda_2}{\sqrt{2}} \,&\, \displaystyle\, \frac{\lambda}{2} \,&\, \displaystyle\, \frac{3\lambda_2}{2} \,&\, \displaystyle\, \frac{\lambda_8}{4} \,&\, \displaystyle\, \frac{\lambda_L}{2} \,&\, \displaystyle\, \frac{\lambda_2}{2} \\
\displaystyle\, \frac{\lambda_7}{2\sqrt{2}} \,&\, \displaystyle\, \frac{\lambda_8}{2\sqrt{2}} \,&\, \displaystyle\, \frac{\lambda_7}{4} \,&\, \displaystyle\, \frac{\lambda_8}{4} \,&\, \displaystyle\, \frac{3\lambda_6}{8} \,&\, \displaystyle\, \frac{\lambda_7}{4} \,&\, \displaystyle\, \frac{\lambda_8}{4} \\
\displaystyle\, \frac{\lambda_1}{\sqrt{2}} \,&\, \displaystyle\, \frac{\lambda_3}{\sqrt{2}} \,&\, \displaystyle\, \frac{\lambda_1}{2} \,&\, \displaystyle\, \frac{\lambda_L}{2} \,&\, \displaystyle\, \frac{\lambda_7}{4} \,&\, \displaystyle\, \frac{3\lambda_1}{2} \,&\, \displaystyle\, \frac{\lambda}{2} \\
\displaystyle\, \frac{\lambda_3}{\sqrt{2}} \,&\, \displaystyle\, \frac{\lambda_2}{\sqrt{2}} \,&\, \displaystyle\, \frac{\lambda_L}{2} \,&\, \displaystyle\, \frac{\lambda_2}{2} \,&\, \displaystyle\, \frac{\lambda_8}{4} \,&\, \displaystyle\, \frac{\lambda}{2} \,&\, \displaystyle\, \frac{3\lambda_2}{2} \\
\end{array}
\right)\nonumber\\
\end{eqnarray}

Despite its apparently complicated structure, three eigenvalues of  ${\cal M}_2$ are located as roots of the following cubic polynomial equation:
\begin{eqnarray}
&&2\,x^3 + (-6\lambda_1-6\lambda_2-\lambda_6)\,x^2 + (18\lambda_1\lambda_2-8\lambda_3^2-8\lambda_3\lambda_4-\nonumber\\
&&2\lambda_4^2+3\lambda_1\lambda_6+3\lambda_2\lambda_6-\lambda_7^2-\lambda_8^2)\,x -9\lambda_1\lambda_2\lambda_6 + 4\lambda_3^2\lambda_6 +\nonumber\\
&& 4\lambda_3\lambda_4\lambda_6  + \lambda_4^2\lambda_6 + 3\lambda_2\lambda_7^2  - 4\lambda_3\lambda_7\lambda_8 - \nonumber\\
&& 2\lambda_4\lambda_7\lambda_8 + 3\lambda_1\lambda_8^2 = 0
\end{eqnarray}
Those roots are denoted as $b_1, b_2$ and $b_3$. The remaining five eigenvalues of ${\cal M}_2$ are as follows:
\begin{eqnarray}
& & b_{4} = \frac{\lambda_6}{4} \\
& & b_{\pm}= \frac{1}{2}(\lambda_1+\lambda_2\pm\sqrt{(\lambda_1-\lambda_2)^2+4\lambda_4^2}) \\ 
& & f_{\pm}= \frac{1}{2}(\lambda_1+\lambda_2\pm\sqrt{(\lambda_1-\lambda_2)^2+4\lambda_5^2})
\end{eqnarray}
The second submatrix ${\cal M}_3$ corresponds to scattering with one of the following
initial and final states:
$(\phi_1\chi_1$, $\phi_2\chi_2)$. One finds that ${\cal M}_3$  is given by

\begin{eqnarray}
{\cal M}_3=\left(
\begin{array}{cc}
\displaystyle \lambda_1 & \lambda_5  \\
\displaystyle \lambda_5 & \lambda_2  \\
\end{array}
\right)
\end{eqnarray}
$\ $\\
the 3 eigenvalues read as follows:
\begin{eqnarray}
& & c_{\pm} = f_{\pm} \ \ 
\end{eqnarray}
The fourth submatrix ${\cal M}_4$ corresponds to scattering with
initial and final states being one of the following $12$ sates :
($\phi_1\phi_1^+$,$\chi_1\phi_1^+$,$\phi_2\phi_1^+$,$\chi_2\phi_1^+$,$\phi_s\phi_1^+$,$\phi_1\phi_2^+$,$\chi_1\phi_2^+$,$\phi_2\phi_2^+$,$\chi_2\phi_2^+$,$\phi_s\phi_2^+$). It reads, 

\begin{widetext}
\begin{eqnarray}
{\cal M}_4 = \left(
\begin{array}{cccccccccccc}
\displaystyle\, \lambda_1 \,&\, \displaystyle\, 0 \,&\, \displaystyle\, 0 \,&\, \displaystyle\, 0 \,&\, \displaystyle\, 0 \,&\, \displaystyle\, 0 \,&\, \displaystyle\, 0 \,&\, \displaystyle\, \frac{\lambda_{45}^+}{2} \,&\, \displaystyle\, -i\frac{\lambda_{45}^-}{2} \,&\, \displaystyle\, 0  \\

\displaystyle\, 0 \,&\, \displaystyle\, \lambda_1 \,&\, \displaystyle\, 0 \,&\, \displaystyle\, 0 \,&\, \displaystyle\, 0 \,&\, \displaystyle\, 0 \,&\, \displaystyle\, 0 \,&\, \displaystyle\, i\frac{\lambda_{45}^-}{2} \,&\, \displaystyle\, \frac{\lambda_{45}^+}{2} \,&\, \displaystyle\, 0  \\

\displaystyle\, 0 \,&\, \displaystyle\, 0 \,&\, \displaystyle\, \lambda_3 \,&\, \displaystyle\, 0 \,&\, \displaystyle\, 0  \,&\, \displaystyle\, \frac{\lambda_{45}^+}{2} \,&\, \displaystyle\, i\frac{\lambda_{45}^-}{2} \,&\, \displaystyle\, 0 \,&\, \displaystyle\, 0 \,&\, \displaystyle\, 0  \\

\displaystyle\, 0 \,&\, \displaystyle\, 0 \,&\, \displaystyle\, 0 \,&\, \displaystyle\, \lambda_3 \,&\, \displaystyle\, 0 \,&\, \displaystyle\, -i\frac{\lambda_{45}^-}{2} \,&\, \displaystyle\, \frac{\lambda_{45}^+}{2} \,&\, \displaystyle\, 0 \,&\, \displaystyle\, 0 \,&\, \displaystyle\, 0  \\

\displaystyle\, 0 \,&\, \displaystyle\, 0 \,&\, \displaystyle\, 0 \,&\, \displaystyle\, 0 \,&\, \displaystyle\, \frac{\lambda_7}{2} \,&\, \displaystyle\, 0 \,&\, \displaystyle\, 0 \,&\, \displaystyle\, 0 \,&\, \displaystyle\, 0 \,&\, \displaystyle\, 0  \\

\displaystyle\, 0 \,&\, \displaystyle\, 0 \,&\, \displaystyle\, \frac{\lambda_{45}^+}{2} \,&\, \displaystyle\, i\frac{\lambda_{45}^-}{2} \,&\, \displaystyle\, 0 \,&\, \displaystyle\, \lambda_3 \,&\, \displaystyle\, 0 \,&\, \displaystyle\, 0 \,&\, \displaystyle\, 0 \,&\, \displaystyle\, 0  \\

\displaystyle\, 0 \,&\, \displaystyle\, 0 \,&\, \displaystyle\, -i\frac{\lambda_{45}^-}{2} \,&\, \displaystyle\, \frac{\lambda_{45}^+}{2} \,&\, \displaystyle\, 0  \,&\, \displaystyle\, 0 \,&\, \displaystyle\, \lambda_3 \,&\, \displaystyle\, 0 \,&\, \displaystyle\, 0 \,&\, \displaystyle\, 0  \\

\displaystyle\, \frac{\lambda_{45}^+}{2} \,&\, \displaystyle\, -i\frac{\lambda_{45}^-}{2} \,&\, \displaystyle\, 0 \,&\, \displaystyle\, 0 \,&\, \displaystyle\, 0 \,&\, \displaystyle\, 0 \,&\, \displaystyle\, 0 \,&\, \displaystyle\, \lambda_2 \,&\, \displaystyle\, 0 \,&\, \displaystyle\, 0  \\

\displaystyle\, i\frac{\lambda_{45}^-}{2} \,&\, \displaystyle\, \frac{\lambda_{45}^+}{2} \,&\, \displaystyle\, 0 \,&\, \displaystyle\, 0 \,&\, \displaystyle\, 0 \,&\, \displaystyle\, 0 \,&\, \displaystyle\, 0 \,&\, \displaystyle\, 0 \,&\, \displaystyle\, \lambda_2 \,&\, \displaystyle\, 0  \\

\displaystyle\, 0 \,&\, \displaystyle\, 0 \,&\, \displaystyle\, 0 \,&\, \displaystyle\, 0 \,&\, \displaystyle\, 0 \,&\, \displaystyle\, 0 \,&\, \displaystyle\, 0 \,&\, \displaystyle\, 0 \,&\, \displaystyle\, 0 \,&\, \displaystyle\, \frac{\lambda_8}{2}  \\
\end{array}
\right)\nonumber\\
\end{eqnarray}
\end{widetext}

The corresponding eigenvalues are :
\begin{eqnarray}
& & d_{1} =  a_1,\quad d_{2} =  a_2,\quad d_{3} = a_3,\quad  d_{4} = a_4  \\
& & d_{5} = a_5,\quad d_{\pm} = b_{\pm},\quad g_{\pm} = f_{\pm} \\
& & d_{6} = \lambda_3-\lambda_4
\end{eqnarray}

The fifth submatrix ${\cal M}_5$ corresponds to scattering with
initial and final states being one of the following $3$ sates:
$(\frac{\phi_1^+\phi_1^+}{\sqrt{2}}$,$\frac{\phi_2^+\phi_2^+}{\sqrt{2}}$,$\phi_1^+\phi_2^+)$. It reads, 

\begin{eqnarray}
{\cal M}_5=\left(
\begin{array}{ccc}
\displaystyle\, \lambda_1 \,&\, \displaystyle\, \lambda_5 \,&\, \displaystyle\,  0  \\
\displaystyle\, \lambda_5 &\, \displaystyle\, \lambda_2 &\, \displaystyle\, 0  \\
\displaystyle\, 0 &\, \displaystyle\,  0 &\, \displaystyle\, \lambda_{34}^+ \\
\end{array}
\right)
\end{eqnarray}
$\ $\\
and possesses the following 3 distinct eigenvalues:
\begin{eqnarray}
& & e_{1} = a_1\\
& & e_{\pm} = f_{\pm}
\end{eqnarray}
%

\setcounter{equation}{0}
\renewcommand{\theequation}{C\arabic{equation}}
\section{Oblique Parameters}
\label{Oblic}
In order to study the contribution of the oblique parameters in the framework of 
2HDMS, we have used the general expressions presented in ~\cite{Grimus,Lavoura,G.2003} for 
the $SU(2)\times U(1)$ electroweak model with an arbitrary number of scalar $SU(2)$ doublets, with hypercharge $\pm \frac{1}{2}$, and an arbitrary number of scalar singlets. 

In this study, we have four real fields that are related to the physical fields $h_{1,2,3}$ and $A$ 
through, 
\begin{eqnarray} 
\begin{pmatrix}\varphi_a+i G \\ \varphi_b + iA \\ \varphi_c \end{pmatrix} = 
\begin{pmatrix} i & R_{11} & R_{12} & R_{13} & 0 \\
0 & R_{21} & R_{22} & R_{23} & i \\
0 & R{31} & R_{32} & R_{23} & 0 \end{pmatrix}  \begin{pmatrix} G \\ h_1 \\ h_2 \\ h_3 \\ A\end{pmatrix}
\end{eqnarray}
where $R_{ij}$ are defined in terms of the mixing angle $\beta$ and the elements of the rotation matrix given by eq, $\mathcal{R}_{ij}$, as follows,
\begin{eqnarray} 
R_{11}&=\mathcal{R}_{11} c_\beta + s_\beta \mathcal{R}_{12},\ \ R_{21}=\mathcal{R}_{12} c_\beta-s_\beta \mathcal{R}_{11},\ \ R_{13}=\mathcal{R}_{13}  \nonumber \\
R_{12}&=\mathcal{R}_{21} c_\beta+\mathcal{R}_{22} s_\beta,\ \ R_{22}= \mathcal{R}_{22} c_\beta -s_\beta \mathcal{R}_{21},\ \ R_{32}=\mathcal{R}_{23} \nonumber \\
R_{13}&=c_\beta \mathcal{R}_{31}+s_\beta \mathcal{R}_{32},\ \ R_{23}=\mathcal{R}_{32} c_\beta-s_\beta \mathcal{R}_{31},\ \ R_{33}=\mathcal{R}_{33} \nonumber 
\end{eqnarray}

We recall that the charged sector is the same as 2HDM, It contains only a pair of charged scalars $H^\pm$. As a result the charged field is related to the physical charged scalar field through the unit matrix.

Therefore, the oblique parameters S, T and U in the 2HDMS are given by :
\begin{widetext}
\begin{align}
	S &= \frac{1}{24 \pi}[(2s_W^2-1)^2G(m_{H^+}^2,m_{H^+}^2,m_Z^2)+R_{21}^2 G(m_{H_1}^2,m_{A}^2,m_Z^2)\nonumber  \\
	&+ R_{22}^2 G(m_{H_2}^2,m_{A}^2,m_Z^2)+R_{23}^2 G(m_{H_3}^2,m_{A}^2,m_Z^2)+(R_{11}^2+R_{21}^2)\ln(m_{H_1}^2)  \nonumber \\ 
	&+(R_{12}^2+R_{22}^2)\ln(m_{H_2}^2) + (R_{13}^2 +R_{23}^2)\ln(m_{H_3}^3)+\ln(m_A^2)-2 \ln(m_{H^+}^2)-\ln(m_{h_{ref}}^2) \nonumber \\
	&+ R_{11}^2 \hat G(m_{H_{1}^2},m_Z^2)+R_{12}^2 \hat G(m_{H_{2}}^2,m_Z^2)+R_{13}^2 \hat G(m_{H_{3}}^2,m_Z^2)-\hat G(m_{h_{ref}}^2,m_Z^2)]
\end{align} 
\begin{align}
	T= &\frac{1}{16 \pi^2 m_W^2 s_W^2}(R_{21}^2 F(m_{H^+}^2,m_{H_1}^2)+R_{22}^2 F(m_{H^+}^2,m_{H_2}^2)+R_{23}^2 F(m_{H^+}^2,m_{H_3}^2) \nonumber\\
	& +F(m_{H^+}^2,m_{A}^2)-R_{21}^2 F(m_{H_1}^2,m_{A}^2)-R_{22}^2 F(m_{H_2}^2,m_{A}^2)-R_{23}^2 F(m_{H_3}^2,m_{A}^2)  \nonumber\\
	& +3(R_{11}^2 (F(m_Z^2,m_{H_1}^2)-F(m_W^2,m_{H_1}^2))+R_{12}^2 (F(m_Z^2,m_{H_2}^2)-F(m_W^2,m_{H_2}^2))   \nonumber \\
	&+  R_{13}^2 (F(m_Z^2,m_{H_3}^2)-F(m_W^2,m_{H_3}^2))) - 3(F(m_Z^2,m_{h_{ref}}^2)-F(m_W^2,m_{h_{ref}}^2))) 
\end{align}
and
\begin{align}
	U &= \frac{1}{24 \pi}[R_{21}^2 G(m_{H^+}^2,m_{H_1}^2,m_W^2)+R_{22}^2 G(m_{H^+}^2,m_{H_2}^2,m_W^2)\nonumber \\ 
	&+R_{23}^2 G(m_{H^+}^2,m_{H_3}^2,m_W^2)+ G(m_{H^+}^2,m_{A}^2,m_W^2)-(2s_W^2-1)^2G(m_{H^+}^2,m_{H^+}^2,m_Z^2)\nonumber \\ 
	&-(R_{21}^2 G(m_{H_1}^2,m_{A}^2,m_Z^2)-R_{22}^2 G(m_{H_2}^2,m_{A}^2,m_Z^2)-R_{23}^2 G(m_{H_3}^2,m_{A}^2,m_Z^2)\nonumber \\ 
	&+R_{11}^2(\hat G(m_{H_1}^2,m_W^2)-\hat G(m_{H_1}^2,m_Z^2))+R_{12}^2(\hat G(m_{H_2}^2,m_W^2)\nonumber  \\
	&-\hat G(m_{H_2}^2,m_Z^2))+ R_{13}^2(\hat G(m_{H_3}^2,m_W^2)-\hat G(m_{H_3}^2,m_Z^2))\nonumber  \\
	&-G(m_{h_{ref}}^2,m_W^2)+G(m_{h_{ref}}^2,m_Z^2)] 
\end{align}
\end{widetext}
where $m_{h_{ref}}$ is the reference mass of the neutral SM Higgs. \\

The explicit forms of these functions, $F \left( x, y \right)$, $G \left( I, J, Q \right)$ and $\hat G \left( I, Q \right)$
are given by ~(\ref{F}), ~(\ref{G1}) and~(\ref{G2}).
\begin{equation}
F \left( x, y \right) \equiv
\left\{ \begin{array}{ll}
{\displaystyle \frac{x+y}{2} - \frac{xy}{x-y}\, \ln{\frac{x}{y}}}
&\Leftarrow\ x \neq y,
\\*[3mm]
0 &\Leftarrow\ x = y.
\end{array} \right. 
\label{F}
\end{equation}
\begin{eqnarray}
&&G \left( I, J, Q \right) = - \frac{16}{3} + \frac{5 \left( I + J \right)}{Q} - \frac{2 \left( I - J \right)^2}{Q^2}\nonumber\\ 
&&\hspace{0.5cm}+\frac{3}{Q}\left[ \frac{I^2 + J^2}{I - J}-\frac{I^2 - J^2}{Q}+\frac{\left( I - J \right)^3}{3 Q^2} \right]
\ln{\frac{I}{J}}\nonumber\\
&&\hspace{0.5cm}+\frac{r}{Q^3}\, f \left( t, r \right).
\label{G1}
\end{eqnarray}
If  $I=J$, $G(I,J,Q)$ is :
\begin{eqnarray}
G \left( I, J, Q \right) &=& - \frac{16}{3} +\frac{16}{Q} I
+ \frac{r}{Q^3}\, f \left( t, r \right) \nonumber 
\end{eqnarray}
and :
\begin{eqnarray}
\label{f}
f \left( t, r \right) \equiv \left\{ \begin{array}{lcl}
{\displaystyle
	\sqrt{r}\, \ln{\left| \frac{t - \sqrt{r}}{t + \sqrt{r}} \right|}
} & \Leftarrow & r > 0,
\\*[3mm]
0 & \Leftarrow & r = 0,
\\*[2mm]
{\displaystyle
	2\, \sqrt{-r}\, \arctan{\frac{\sqrt{-r}}{t}}
} & \Leftarrow & r < 0.
\end{array} \right. \nonumber
\end{eqnarray}
with :
\begin{eqnarray}
\label{tr}
t \equiv I + J - Q
\quad \mbox{and} \quad
r \equiv Q^2 - 2 Q \left( I + J \right) + \left( I - J \right)^2 \nonumber
\end{eqnarray}
\begin{eqnarray}
&&\hat G \left( I, Q \right) =- \frac{79}{3} + 9\, \frac{I}{Q} - 2\, \frac{I^2}{Q^2}\nonumber\\
&&+ \bigg( - 10 + 18\, \frac{I}{Q} - 6\, \frac{I^2}{Q^2} + \frac{I^3}{Q^3}
- 9\, \frac{I + Q}{I - Q} \bigg) \ln{\frac{I}{Q}} \nonumber \\ 
&&+ \bigg( 12 - 4\, \frac{I}{Q} + \frac{I^2}{Q^2} \bigg)
\frac{f \left( I, I^2 - 4 I Q \right)}{Q}.
\label{G2}
\end{eqnarray}

\section{Scalar couplings}
\label{scalarcoup}
We list hereafter the triple scalar couplings needed for our study.
\setcounter{equation}{0}
\renewcommand{\theequation}{D\arabic{equation}}
\begin{widetext}
\begin{table}[h!]
\centering
\begin{tabular}{cc} 
\hline
Vertex & Coupling  \\ \hline
$H_1 H^+ H^-$ & $-i \frac{1}{v} \bigg(-\tilde{\mu}^2 (\frac{\mathcal{R}_{11}}{c_\beta}+\frac{\mathcal{R}_{12}}{s_\beta}) + m_{H_1}^2 (\frac{\mathcal{R}_{11} s_\beta^2}{c_\beta}+\frac{\mathcal{R}_{12} c_\beta^2}{s_\beta}) + 2 m_{H^\pm}^2 (\mathcal{R}_{11} c_\beta + \mathcal{R}_{12} s_\beta ) \bigg)$ \\ 
$H_2 H^+ H^-$ & $-i \frac{1}{v} \bigg(-\tilde{\mu}^2 (\frac{\mathcal{R}_{21}}{c_\beta}+\frac{\mathcal{R}_{22}}{s_\beta}) + m_{H_2}^2 (\frac{\mathcal{R}_{21} s_\beta^2}{c_\beta}+\frac{\mathcal{R}_{22} c_\beta^2}{s_\beta}) + 2 m_{H^\pm}^2 (\mathcal{R}_{21} c_\beta + \mathcal{R}_{22} s_\beta ) \bigg)$ \\ 
$H_3 H^+ H^-$ & $-i \frac{1}{v} \bigg(-\tilde{\mu}^2 (\frac{\mathcal{R}_{31}}{c_\beta}+\frac{\mathcal{R}_{32}}{s_\beta}) + m_{H_3}^2 (\frac{\mathcal{R}_{31} s_\beta^2}{c_\beta}+\frac{\mathcal{R}_{32} c_\beta^2}{s_\beta}) + 2 m_{H^\pm}^2 (\mathcal{R}_{31} c_\beta + \mathcal{R}_{32} s_\beta ) \bigg)$ \\
$H_1 H_1 H_1 $ & $-i\frac{3}{v} \bigg( -\tilde{\mu}^2 (\mathcal{R}_{12}^2 c_\beta (\mathcal{R}_{12} \frac{c_\beta}{s_\beta} - \mathcal{R}_{11}) + \mathcal{R}_{11}^2 s_\beta (\mathcal{R}_{11} \frac{s_\beta}{c_\beta} - \mathcal{R}_{12}))+ \frac{m_{H_1}^2}{v_S} ( \mathcal{R}_{13}^3 v + \mathcal{R}_{12}^3 \frac{v_S}{s_\beta} + \mathcal{R}_{11}^3 \frac{v_S}{c_\beta}) \bigg)$ \\ 
$H_2 H_2 H_2 $ & $-i\frac{3}{v} \bigg( -\tilde{\mu}^2 (\mathcal{R}_{22}^2 c_\beta (\mathcal{R}_{22} \frac{c_\beta}{s_\beta} - \mathcal{R}_{21}) + \mathcal{R}_{21}^2 s_\beta (\mathcal{R}_{21} \frac{s_\beta}{c_\beta} - \mathcal{R}_{22}))+ \frac{m_{H_2}^2}{v_S} ( \mathcal{R}_{23}^3 v + \mathcal{R}_{22}^3 \frac{v_S}{s_\beta} + \mathcal{R}_{21}^3 \frac{v_S}{c_\beta}) \bigg)$ \\ 
$H_3 H_3 H_3 $ & $-i\frac{3}{v} \bigg( -\tilde{\mu}^2 (\mathcal{R}_{32}^2 c_\beta (\mathcal{R}_{32} \frac{c_\beta}{s_\beta} - \mathcal{R}_{31}) + \mathcal{R}_{31}^2 s_\beta (\mathcal{R}_{31} \frac{s_\beta}{c_\beta} - \mathcal{R}_{32}))+ \frac{m_{H_3}^2}{v_S} ( \mathcal{R}_{33}^3 v + \mathcal{R}_{32}^3 \frac{v_S}{s_\beta} + \mathcal{R}_{31}^3 \frac{v_S}{c_\beta}) \bigg)$ \\
$H_2 H_1 H_1 $ & $-i\frac{1}{v} \bigg( 
-\frac{1}{2} \tilde{\mu}^2 \big( \frac{\mathcal{R}_{12}}{s_\beta} - \frac{\mathcal{R}_{11}}{c_\beta} \big)  \big( 6 \mathcal{R}_{12} \mathcal{R}_{22} + 6 \mathcal{R}_{13} \mathcal{R}_{23} s_\beta^2 + 2 \mathcal{R}_{33} s_\beta c_\beta \big)    
+ \frac{2 m_{H_1}^2+m_{H_2}^2}{v_S} \big( \mathcal{R}_{13}^2 \mathcal{R}_{23} v + $\\
& $ \mathcal{R}_{12}^2 \mathcal{R}_{22} \frac{v_S}{s_\beta} + \mathcal{R}_{11}^2 \mathcal{R}_{21} \frac{v_S}{c_\beta} \big)
 \bigg)$ \\
$H_3 H_1 H_1 $ & $-i\frac{1}{v} \bigg( 
-\frac{1}{2} \tilde{\mu}^2 \big( \frac{\mathcal{R}_{12}}{s_\beta} - \frac{\mathcal{R}_{11}}{c_\beta} \big)  \big( 6 \mathcal{R}_{12} \mathcal{R}_{32} + 6 \mathcal{R}_{13} \mathcal{R}_{33} s_\beta^2 - 2 \mathcal{R}_{23} s_\beta c_\beta \big)    
+ \frac{2 m_{H_1}^2+m_{H_3}^2}{v_S} \big( \mathcal{R}_{13}^2 \mathcal{R}_{33} v + $\\
& $ \mathcal{R}_{12}^2 \mathcal{R}_{32} \frac{v_S}{s_\beta} + \mathcal{R}_{11}^2 \mathcal{R}_{31} \frac{v_S}{c_\beta} \big)
 \bigg)$ \\
$H_3 H_2 H_2 $ & $-i\frac{1}{v} \bigg( 
-\frac{1}{2} \tilde{\mu}^2 \big( \frac{\mathcal{R}_{22}}{s_\beta} - \frac{\mathcal{R}_{21}}{c_\beta} \big)  \big( 6 \mathcal{R}_{22} \mathcal{R}_{32} + 6 \mathcal{R}_{23} \mathcal{R}_{33} s_\beta^2 + 2 \mathcal{R}_{13} s_\beta c_\beta \big)    
+ \frac{2 m_{H_2}^2+m_{H_3}^2}{v_S} \big( \mathcal{R}_{23}^2 \mathcal{R}_{33} v + $\\
& $ \mathcal{R}_{22}^2 \mathcal{R}_{32} \frac{v_S}{s_\beta} + \mathcal{R}_{21}^2 \mathcal{R}_{31} \frac{v_S}{c_\beta} \big)
 \bigg)$ \\
\hline
\end{tabular}
\caption{Trilinear Higgs boson self-interactions ($i\lambda 3H$) in the Feynman gauge within the 2HDMS.} 
\label{tri}
\end{table}
\end{widetext}
\noindent
with $\tilde{\mu}^2 = \frac{\mu^2}{s_\beta c_\beta}$.

\bibliographystyle{unsrt}


\begin{thebibliography}{99}


\bibitem{atlasdiscovery}
G.~Aad {\it et al.} [ATLAS Collaboration],
Phys.\ Lett.\ B {\bf 716}, 1 (2012)
doi:10.1016/j.physletb.2012.08.020
[arXiv:1207.7214 [hep-ex]].


\bibitem{cmsdiscovery}
S.~Chatrchyan {\it et al.} [CMS Collaboration],
  Phys.\ Lett.\ B {\bf 716}, 30 (2012)
  doi:10.1016/j.physletb.2012.08.021
  [arXiv:1207.7235 [hep-ex]].





\bibitem{Sirunyan:2018hoz} 
  A.~M.~Sirunyan {\it et al.} [CMS Collaboration],
  Phys.\ Rev.\ Lett.\  {\bf 120}, no. 23, 231801 (2018)
  doi:10.1103/PhysRevLett.120.231801, 10.1130/PhysRevLett.120.231801
  [arXiv:1804.02610 [hep-ex]].



\bibitem{Aaboud:2018urx} 
  M.~Aaboud {\it et al.} [ATLAS Collaboration],
  arXiv:1806.00425 [hep-ex].
  
\bibitem{accuracy1}  
  S.~Dawson {\it et al.}, 
  arXiv:1310.8361 [hep-ex];
  D.~Zeppenfeld, R.~Kinnunen, A.~Nikitenko and E.~Richter-Was,
  Phys.\ Rev.\ D {\bf 62} (2000) 013009
  [hep-ph/0002036];
F.~Gianotti and M.~Pepe-Altarelli,
  Nucl.\ Phys.\ Proc.\ Suppl.\  {\bf 89}, 177 (2000)
  doi:10.1016/S0920-5632(00)00841-0
  [hep-ex/0006016].


\bibitem{accuracy2}  
C.~Englert, A.~Freitas, M.~M.~Muhlleitner, T.~Plehn, M.~Rauch, M.~Spira and K.~Walz,
  J.\ Phys.\ G {\bf 41}, 113001 (2014)
  doi:10.1088/0954-3899/41/11/113001
  [arXiv:1403.7191 [hep-ph]].


  \bibitem{accuracy3}  
G.~Moortgat-Pick {\it et al.},
  Eur.\ Phys.\ J.\ C {\bf 75}, no. 8, 371 (2015)
  doi:10.1140/epjc/s10052-015-3511-9
  [arXiv:1504.01726 [hep-ph]].
  
  \bibitem{Robens:2015gla} 
  T.~Robens and T.~Stefaniak,
  Eur.\ Phys.\ J.\ C {\bf 75}, 104 (2015)
  doi:10.1140/epjc/s10052-015-3323-y
  [arXiv:1501.02234 [hep-ph]].
    
 
  \bibitem{Bernon:2015qea} 
  J.~Bernon, J.~F.~Gunion, H.~E.~Haber, Y.~Jiang and S.~Kraml,
  Phys.\ Rev.\ D {\bf 92}, no. 7, 075004 (2015)
  doi:10.1103/PhysRevD.92.075004
  [arXiv:1507.00933 [hep-ph]].
  
 
  \bibitem{Ma:2006km} 
  E.~Ma,
  Phys.\ Rev.\ D {\bf 73}, 077301 (2006)
  doi:10.1103/PhysRevD.73.077301
  [hep-ph/0601225].
  
  \bibitem{Arhrib:2013ela} 
  A.~Arhrib, Y.~L.~S.~Tsai, Q.~Yuan and T.~C.~Yuan,
  JCAP {\bf 1406}, 030 (2014)
  doi:10.1088/1475-7516/2014/06/030
  [arXiv:1310.0358 [hep-ph]].


\bibitem{Drozd:2014yla} 
  A.~Drozd, B.~Grzadkowski, J.~F.~Gunion and Y.~Jiang,
  JHEP {\bf 1411}, 105 (2014)
  doi:10.1007/JHEP11(2014)105
  [arXiv:1408.2106 [hep-ph]].
      
  \bibitem{Grzadkowski:2009iz} 
  B.~Grzadkowski and P.~Osland,
  Phys.\ Rev.\ D {\bf 82}, 125026 (2010)
  doi:10.1103/PhysRevD.82.125026
  [arXiv:0910.4068 [hep-ph]].
     
      
\bibitem{Chen:2013jvg} 
  C.~Y.~Chen, M.~Freid and M.~Sher,
  Phys.\ Rev.\ D {\bf 89}, no. 7, 075009 (2014)
  doi:10.1103/PhysRevD.89.075009
  [arXiv:1312.3949 [hep-ph]].




\bibitem{Muhlleitner:2016mzt} 
  M.~Muhlleitner, M.~O.~P.~Sampaio, R.~Santos and J.~Wittbrodt,
  JHEP {\bf 1703}, 094 (2017)
  doi:10.1007/JHEP03(2017)094
  [arXiv:1612.01309 [hep-ph]].
  
  

\bibitem{weinberg}  
S.~L.~Glashow and S.~Weinberg,
  Phys.\ Rev.\ D {\bf 15}, 1958 (1977).
  doi:10.1103/PhysRevD.15.1958
  
  


\bibitem{Branco}
G.~C.~Branco, P.~M.~Ferreira, L.~Lavoura, M.~N.~Rebelo, M.~Sher and J.~P.~Silva,
  Phys.\ Rept.\  {\bf 516}, 1 (2012)
  doi:10.1016/j.physrep.2012.02.002
  [arXiv:1106.0034 [hep-ph]].
  
  
  
\bibitem{Gunion:1990kf} 
  J.~F.~Gunion, H.~E.~Haber and J.~Wudka,
  Phys.\ Rev.\ D {\bf 43}, 904 (1991).
  doi:10.1103/PhysRevD.43.904
  
  
  
 \bibitem{Bento:2017eti} 
  M.~P.~Bento, H.~E.~Haber, J.~C.~Romão and J.~P.~Silva,
  JHEP {\bf 1711}, 095 (2017)
  doi:10.1007/JHEP11(2017)095
  [arXiv:1708.09408 [hep-ph]].
  
  
  
 \bibitem{Bento:2018fmy} 
  M.~P.~Bento, H.~E.~Haber, J.~C.~Romao and J.~P.~Silva,
  arXiv:1808.07123 [hep-ph].
  
   
  \bibitem{Kanemura:1993hm} 
  S.~Kanemura, T.~Kubota and E.~Takasugi,
  Phys.\ Lett.\ B {\bf 313}, 155 (1993)
  doi:10.1016/0370-2693(93)91205-2
  [hep-ph/9303263].

   
\bibitem{arhrib}
A.~G.~Akeroyd, A.~Arhrib and E.~M.~Naimi,
  Phys.\ Lett.\ B {\bf 490}, 119 (2000)
  doi:10.1016/S0370-2693(00)00962-X
  [hep-ph/0006035].
and 
A.~Arhrib,
  hep-ph/0012353.
  

\bibitem{Peskin}
M.E. Peskin and T. Takeuchi, A new constraint on a strongly interacting Higgs sector, Phys. Rev. Lett. 65 (1990) 964;
M.E. Peskin and T. Takeuchi, Estimation of oblique electroweak corrections, Phys. Rev. D 46 (1992) 381.


\bibitem{Grimus}
W.~Grimus, L.~Lavoura, O.M.~Ogreid and P.~Osland,
  J.Phys. G35 (2008) 075001
  DOI: 10.1088/0954-3899/35/7/075001, 
  [arXiv:0711.4022v2 [hep-ph]]

 
\bibitem{Lavoura}
W.~Grimus, L.~Lavoura, O.M.~Ogreid and P.~Osland,
  Nucl.Phys. B801 (2008) 81-96,
  DOI: 10.1016/j.nuclphysb.2008.04.019,
  [arXiv:0802.4353 [hep-ph]]	
		
 
\bibitem{G.2003}
John F.~Gunion, Howard E.~Haber,
  Phys.\ Rev.\ D67:075019,2003
  DOI: 10.1103/PhysRevD.67.075019 
  [arXiv:hep-ph/0207010]



\bibitem{Haller 2018}
Johannes Haller, Andreas Hoecker, Roman Kogler, Klaus Mönig, Thomas Peiffer and J$\ddot{o}$rg Stelzer,
  Eur.Phys.J. C78 (2018) no.8, 675,
  DOI: 10.1140/epjc/s10052-018-6131-3,
  [arXiv:1803.01853 [hep-ph]]

  
\bibitem{Su:2001}  
H.~-J.~ He, N.~Polonsky, S.~Su,
  Phys.\ Rev.\ D64:053004,2001
  DOI: 10.1103/PhysRevD.64.053004
  [arXiv:hep-ph/0102144]
    
 
\bibitem{Kanemura.2011}
S.~Kanemura, Y.~Okada, H.~Taniguchi, K.~Tsumura,
  Phys.Lett. B (2011) 09.035,
  DOI: 10.1016/j.physletb.2011.09.035 
  [arXiv:1108.3297] 

  
\bibitem{Amhis}
  Y.~Amhis {\it et al.} [HFLAV Collaboration],
  Eur.\ Phys.\ J.\ C {\bf 77}, no. 12, 895 (2017)
  doi:10.1140/epjc/s10052-017-5058-4
  [arXiv:1612.07233 [hep-ex]].
  
  
  				
\bibitem{misiak}
M.~Misiak and M.~Steinhauser,
  Eur.\ Phys.\ J.\ C {\bf 77} (2017) no.3,  201
  doi:10.1140/epjc/s10052-017-4776-y
  [arXiv:1702.04571 [hep-ph]].

 
\bibitem{Khachatryan:2016whc}
  V.~Khachatryan {\it et al.} [CMS Collaboration],
  JHEP {\bf 1702} (2017) 135
  doi:10.1007/JHEP02(2017)135
  [arXiv:1610.09218 [hep-ex]].

 
\bibitem{Aaboud:2017bja}
  M.~Aaboud {\it et al.} [ATLAS Collaboration],
  Phys.\ Lett.\ B {\bf 776} (2018) 318
  doi:10.1016/j.physletb.2017.11.049
  [arXiv:1708.09624 [hep-ex]].
  
  


\bibitem{rui2017}
  M.~Muhlleitner, M.~O.~P.~Sampaio, R.~Santos and J.~Wittbrodt,
  JHEP {\bf 1703}, 094 (2017)
  doi:10.1007/JHEP03(2017)094
  [arXiv:1612.01309 [hep-ph]].

 
 
\bibitem{Engeln:2018mbg} 
  I.~Engeln, M.~Mühlleitner and J.~Wittbrodt,
  arXiv:1805.00966 [hep-ph].
  
  

\bibitem{Bernon:2015wef} 
  J.~Bernon, J.~F.~Gunion, H.~E.~Haber, Y.~Jiang and S.~Kraml,
  Phys.\ Rev.\ D {\bf 93}, no. 3, 035027 (2016)
  doi:10.1103/PhysRevD.93.035027
  [arXiv:1511.03682 [hep-ph]].
 
   
\bibitem{Arhrib:2017uon} 
  A.~Arhrib, R.~Benbrik, S.~Moretti, A.~Rouchad, Q.~S.~Yan and X.~Zhang,
  JHEP {\bf 1807}, 007 (2018)
  doi:10.1007/JHEP07(2018)007
  [arXiv:1712.05332 [hep-ph]].
   

  
\bibitem{Aad:2015bua} 
G.~Aad {\it et al.} [ATLAS Collaboration],
Eur.\ Phys.\ J.\ C {\bf 76}, no. 4, 210 (2016)
doi:10.1140/epjc/s10052-016-4034-8
[arXiv:1509.05051 [hep-ex]].
  
  
\end{thebibliography}

\end{document}